\documentclass[onecolumn,draftclsnofoot,12pt]{IEEEtran}

\usepackage{enumerate}
\usepackage{amsmath,amsthm}
\usepackage{mathtools}
\usepackage{algorithm,algorithmic}
\usepackage{float}
\usepackage{hyperref}
\usepackage{color}
\usepackage{makeidx}
\usepackage{bbm}
\usepackage{graphicx}
\usepackage{lipsum}
\usepackage{soul}
\usepackage{tabularx}
\usepackage{dsfont}
\usepackage[table,xcdraw]{xcolor}

\usepackage{amsfonts}
\usepackage{times}
\usepackage{graphicx}
\usepackage{latexsym}
\usepackage{dsfont}
\usepackage{amssymb}
\usepackage{amsmath}
\usepackage{cite}
\usepackage{verbatim}
\usepackage{subfigure}

\def\bb0{{\mathbb{0}}}

\def\bb{{\mathbf{b}}}

\def\bw{{\mathbf{w}}}

\def\b0{{\mathbf{0}}}

\def\sf0{{\mathsf{0}}}

\usepackage{epstopdf}

\newcommand{\sref}[1]{{Section}~\ref{#1}}
\newcommand{\fref}[1]{{Fig.}~\ref{#1}}
\newcommand{\aref}[1]{{Algorithm}~\ref{#1}}

\DeclareMathOperator*{\argmax}{arg\,max}

\def\rm{\mathrm}

\newcommand{\subto}{\operatorname{s.t.}}
\newcommand{\argmin}{\operatornamewithlimits{arg\min}}

\begin{document}
\title{Reinforcement Learning of Beam Codebooks in \\ Millimeter Wave and Terahertz MIMO Systems}
\author{Yu Zhang, Muhammad Alrabeiah, and Ahmed Alkhateeb \thanks{Yu Zhang, Muhammad Alrabeiah and Ahmed Alkhateeb are with Arizona State University (Email: y.zhang, malrabei, alkhateeb@asu.edu). This work is supported by the National Science Foundation under Grant No. 1923676. Part of this work was presented at IEEE Asilomar \cite{zhang2021reinforcement}. }}
\maketitle

\begin{abstract}
Millimeter wave (mmWave) and terahertz MIMO systems rely on pre-defined beamforming codebooks for both initial access and data transmission. Being pre-defined, however, those codebooks are commonly not optimized for specific environments, user distributions, and/or possible hardware impairments. This leads to large codebook sizes with high beam training overhead which increases the initial access/tracking latency and makes it hard for these systems to support highly mobile applications. To overcome these limitations, this paper develops a deep reinforcement learning framework that learns how to iteratively optimize the codebook beam patterns (shapes) relying only on the receive power measurements and without requiring any explicit channel knowledge. The developed model learns how to autonomously adapt the beam patterns to best match the surrounding environment, user distribution,  hardware impairments, and array geometry. Further, this approach does not require any knowledge about the channel, array geometry, RF hardware, or user positions. To reduce the learning time, the proposed model designs a novel \emph{Wolpertinger}-variant architecture that is capable of efficiently searching for an optimal policy in a large discrete action space, which is important for large antenna arrays with quantized phase shifters. This complex-valued neural network architecture design respects the practical RF hardware constraints such as the constant-modulus and quantized phase shifter constraints. Simulation results based on the publicly available DeepMIMO dataset confirm the ability of the developed framework to learn near-optimal beam patterns for both line-of-sight (LOS) and non-LOS scenarios and for arrays with hardware impairments without requiring any channel knowledge.
\end{abstract}

\section{Introduction} \label{intro}
Millimeter wave (mmWave) and terahertz (THz) MIMO systems adopt large antenna arrays to compensate for the significant path loss and ensure sufficient receive signal power. Because of the high cost and power consumption of the mixed-circuit components, however, these systems normally rely either fully or partially  on analog beamforming, where transceivers employ networks of phase shifters \cite{Alkhateeb2014MIMO,Alkhateeb2014}. This makes the basic MIMO signal processing functions, such as channel estimation and beamforming design, challenging as the channels are  seen only through the RF lens. This motivates mmWave/THz massive MIMO systems to rely on pre-defined \textit{beamforming codebooks} for both initial access and data transmission \cite{giordani2019,11ad,Alr_vision}. The classical pre-defined beamforming/beamsteering codebooks normally consist of a large number of single-lobe beams, each of which can steer the signal towards one direction. These classical codebooks, though, have several drawbacks: (i) To cover all the possible directions, these codebooks consist of a large number of beams, which makes the search over them associated with high beam training overhead. (ii) The second issue is a blowback from the directivity blessing; classical beamsteering codebooks employ single-lobe beams to maximize directivity, which, in many cases, may not be optimal, especially for Non-Line-of-Sight (NLOS) users.  (iii) Further, the design of the classical codebooks normally assume that the array is calibrated and its geometry is known, which associates this design processing with high cost (due to the need for expensive calibration) and makes it hard to adapt to systems with unknown or arbitrary array geometries.

Another look at the aforementioned drawbacks reveals that they stem from the lack of environment and hardware adaptability. A mmWave/THz system that has a sense of its environment could discover the most frequent signal directions (for both single and multi-path cases) and accordingly tailor its codebook beam \textit{patterns} (directions, shapes, number of lobes, etc.). Furthermore, the system can also overcome the challenges with intrinsic hardware impairments or unknown/arbitrary array geometries by learning how to calibrate its beams to adapt to the given hardware. All that awareness and adaptability can potentially be achieved if the mmWave/THz system incorporates a data-driven and artificially-intelligent component. Towards this goal, leveraging machine learning tools, especially reinforcement learning, is particularly promising. Reinforcement learning models can efficiently learn from the observed data and responses obtained from both the  hardware and environment, which may potentially reduce the channel knowledge requirements. \textbf{Hence, the focus of this work is on developing a reinforcement learning based approach that learns how to adapt the codebook beams to the environment and hardware without requiring explicit channel knowledge}.

\subsection{Prior Work:}

Designing efficient beamforming and combining  is essential for realizing the  potential of MIMO communications, and it has been an important research topic in the literature of MIMO signal processing \cite{Lo1999,Love2003, Li2017, ElAyach2014, Alkhateeb2014MIMO}.  For MIMO systems with no hardware constraints, i.e., with fully-digital processing and no constraints on the RF hardware, maximum ratio transmission and combining maximize the achievable SNR with single-stream transmission/reception  \cite{Lo1999}. To realize these solutions, however, the MIMO system should be able to control the magnitude and phase of the signal at each antennas. When only the phase can be controlled, equal-gain transmission solutions have been developed to maximize the SNR or diversity gains \cite{Love2003}. This is particularly interesting for mmWave and terahertz systems where the beamforming/precoding processing is fully or partially done in the RF domain using analog phase shifters \cite{Alkhateeb2014MIMO}. In these systems, however, the phase shifters can normally take only quantized phase shift values. This associates the search over the large space of quantized phase shift values with high complexity (e.g., for a 32-element antenna array with 2-bit phase shifters,  there are $4^{32}$ possible beamforming vectors) \cite{ Li2017, ElAyach2014, Alkhateeb2014MIMO}. Further, in analog beamforming architectures,  the channel is seen through the RF lens, which makes it hard to acquire at the baseband, especially for systems with arbitrary or unknown array geometries. To address these challenges, the first objective of this paper is to design a reinforcement learning based approach to efficiently learn the analog beamforming patterns that adapt to the surrounding environment and the adopted hardware/array geometry  without requiring explicit channel knowledge.

Given the high complexity/training overhead associated with the beamforming design for quantized analog/hybrid architectures, these architectures normally rely on using pre-defined beam codebooks \cite{Alkhateeb2014,Hur2013,Wang2009,Alkhateeb2016d}.  These classical codebooks, however, are normally  designed to have a large number of narrow beams; each beam has a single lobe and points to one direction. This has several limitations: (i) The large number of beams in these classical codebooks leads to large beam training overhead, which makes it hard for these mmWave/terahertz systems to support highly-mobile applications, (ii) single-lobe beams may not be optimal in scenarios with more than one dominant path such as NLOS situations,  (iii) classical codebook design approaches normally rely on the knowledge of the channels, which is hard to achieve in architectures with analog beamforming, especially if these systems adopt imperfect hardware or deploy arrays with unknown/arbitrary array geometries, and (iv) the classical codebooks typically assume fully-calibrated arrays, which is an expensive process. All the aforementioned limitations have motivated the search for adaptive codebook designing methods that leverage machine learning tools to adapt the learned beams based on the environment and hardware. In \cite{Alrabeiah2020Neural}, the beam codebook learning approaches based on neural networks were proposed and shown to achieve good gain compared to the classical beam steering codebooks. The solutions in \cite{Alrabeiah2020Neural}, though, still require full or partial channel knowledge, which is not simple to obtain in mmWave/THz systems. This motivates the development of environment and hardware aware codebook learning approach that does not require explicit channel knowledge, which is the main focus of this paper.

\subsection{Contribution:}
Developing environment and hardware awareness using machine learning is not straightforward when the mmWave system constraints are considered, e.g., channels are not available, phase shifters have finite and limited resolution, and hardware envelops unknown impairments. In this paper, we develop a deep reinforcement learning based framework that can efficiently learn mmWave beam codebooks while addressing all these challenges. The main contributions of this paper can be summarized as follows:
\begin{itemize}
	\item Designing a deep reinforcement learning based framework that can learn how to optimize the beam pattern for a set of users with similar channels. \textbf{The developed  framework relies only on receive power measurements and does not require any channel knowledge}. This framework adapts the beam pattern based on the surrounding environment and learns how to compensate for the hardware impairments. This is done by utilizing a novel Wolpertinger architecture \cite{Dulacarnold2015} which is designed to efficiently explore the large discrete action space. Further, the proposed model accounts for key hardware constraints such as the phase-only, constant-modulus, and quantized-angle constraints \cite{Alkhateeb2014MIMO}.

    \item Developing a reinforcement learning framework that is capable of learning a codebook of beam patterns optimized to serve the users in the surrounding environment. \textbf{The proposed framework autonomously optimizes the codebook beam patterns based on the environment, user distribution, hardware impairments, and array geometry}. Further, it relies only on the receive power measurements, does not require any position or channel knowledge (which relaxes the synchronization/coherence requirements), and  does not require the users to be stationary during the learning process. This is achieved by developing a novel pre-processing approach that relies on SNR-based feature matrices to  partition/assign the users into clusters based on which parallel neural networks are trained.
	
	\item Extensively evaluating the performance of the proposed codebook learning approaches based on the publicly-available DeepMIMO dataset \cite{DeepMIMO}. These experiments adopt both outdoor and indoor wireless communication scenarios and learn codebooks with different sizes. Further, this evaluation is done both for perfect uniform arrays and for arrays with arbitrary geometries and hardware impairments. These experiments provide a comprehensive evaluation of the proposed reinforcement learning based codebook learning approach.
\end{itemize}

The simulation results show that the proposed approach is capable of learning optimized beam patterns and beam codebooks without the need of providing any channel state information. Instead, based solely on the receive combining gains, the deep reinforcement learning solution adjusts the phases of the beamforming vectors to increase the receive gain and finally yield significant improvements over the classical beamsteering codebooks.
\section{System and Channel Models} \label{sec:System}

In this section, we introduce in detail our adopted system and channel models. We also describe how the model considers arbitrary arrays with possible hardware impairments.

\subsection{System Model}

\begin{figure}[t]
	\includegraphics[width=\linewidth]{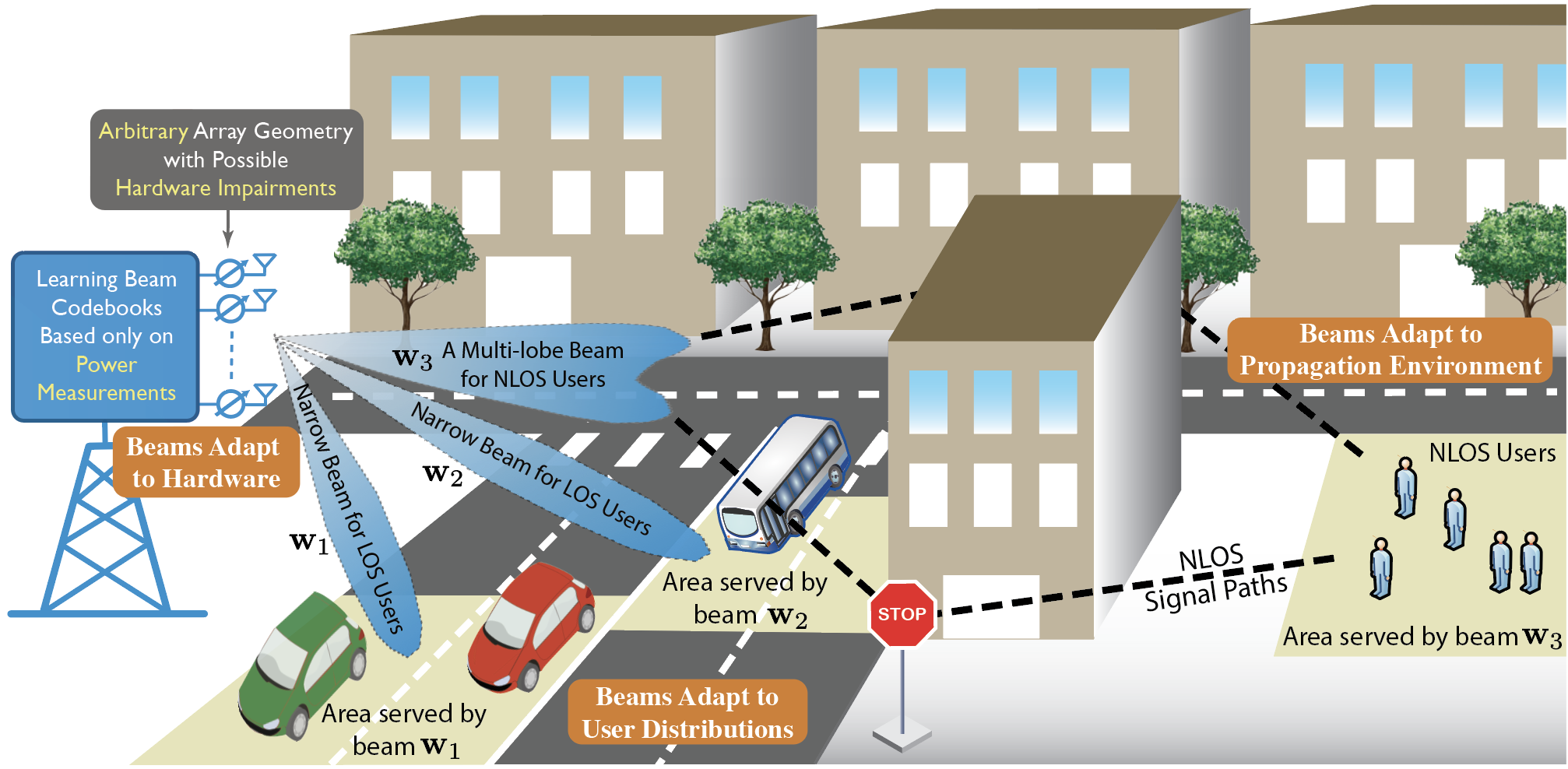}
	\caption{A mmWave/massive MIMO base station with an arbitrary antenna array serving users with a beam codebook $\boldsymbol{\mathcal{W}}$. The objective is to develop a learning approach for adapting the codebook $\boldsymbol{\mathcal{W}}$ to match the given hardware and environment based only on SNR measurements (which relaxes the coherence/synchronization requirements).}
	\label{Sys_model}
\end{figure}

We consider the system model shown in \fref{Sys_model} where a mmWave massive MIMO base station with $M$ antennas is communicating with a single-antenna user. Further, given the high cost and power consumption of mixed-signal components \cite{Alkhateeb2014,HeathJr2016}, we consider a practical system where the base station has only one radio frequency (RF) chain and employs analog-only beamforming using a network of $r$-bit quantized phase shifters. To facilitate the system operation and to respect the hardware constraints, mmWave and massive MIMO systems typically use beamforming codebooks in serving their users. Let $\boldsymbol{\mathcal{W}}$ denote the beam codebook adopted by the base station and assume that it contains $N$ beamforming/combining vectors, with each one of them taking the form
\begin{equation}\label{Analog}
  {\bf w} = \frac{1}{\sqrt{M}}\left[ e^{j\theta_1}, e^{j\theta_2}, \dots, e^{j\theta_M} \right]^T,
\end{equation}
where each phase shift $\theta_m$ is selected from a finite set $\boldsymbol{\Theta}$ with $2^r$ possible discrete values drawn uniformly from $(-\pi, \pi]$.
In the uplink transmission, if a user $u$ transmits a symbol $x\in\mathbb{C}$ to the base station, where the transmitted symbol satisfies the average power constraint $\mathbb{E}\left[|x|^2\right]=P_x$, the received signal at the base station after combining can be expressed as
\begin{equation}\label{sys}
  y_u = {\mathbf w}^H{\mathbf h}_ux + {\mathbf w}^H{\mathbf n},
\end{equation}
where ${\mathbf h}_u\in\mathbb{C}^{M\times 1}$ is the uplink channel vector between the user $u$ and the base station antennas and ${\mathbf n}\sim\mathcal{N}_\mathbb{C}\left(0, \sigma_n^2{\bf I}\right)$ is the receive noise vector at the base station.

\subsection{Channel Model} \label{subsec:channel}

We adopt a general geometric channel model for ${\mathbf h}_u$ \cite{HeathJr2016,Alrabeiah2019}. Assume that the signal propagation between the user $u$ and the base station consists of $L$ paths. Each path $\ell$ has a complex gain $\alpha_\ell$ and an angle of arrival $\phi_\ell$. Then, the channel vector can be written as
\begin{equation}\label{channel}
  {\mathbf h}_u = \sum\limits_{\ell=1}^{L}\alpha_\ell{\mathbf a}(\phi_\ell),
\end{equation}
where ${\bf a}(\phi_\ell)$ is the array response vector of the base station. The definition of ${\bf a}(\phi_\ell)$ depends on the array geometry and hardware impairments. Next, we discuss that in more detail.

\subsection{Hardware Impairments Model} \label{subsec:impairments}

Most of the prior work on mmWave signal processing has assumed uniform antenna arrays with perfect calibration and ideal hardware \cite{Hur2013,Wang2009,Alkhateeb2014,Alkhateeb2014MIMO}. In this paper, we consider a more general antenna array model that accounts for arbitrary geometry and hardware imperfections, and target learning efficient beam codebooks for these systems. This is very important for several reasons: (i) there are scenarios where designing arbitrary arrays is needed, for example, to improve the angular resolution or enhance the direction-of-arrival estimation performance \cite{Pal2010,Rubsamen2009}, (ii) the fabrication process of large mmWave arrays normally has some imperfections, and (iii) the calibration process of the  mmWave phased arrays is an expensive process that requires special high-performance RF circuits \cite{Moon2019}.
While the codebook learning solutions that we develop in this paper are general for various kinds of arrays and hardware impairments, we evaluate them in \sref{sec:Results} with respect to two main characteristics of interest, namely  non-uniform spacing and phase mismatch between the antenna elements.  For linear arrays, the array response vector can be modeled to capture these characteristics as follows
\begin{equation}\label{ARV-cor}
	{{\bf a}}(\phi_\ell) = \left[ e^{j\left(kd_1\cos(\phi_\ell) + \Delta\theta_1\right)}, e^{j\left(kd_2\cos(\phi_\ell) +\Delta\theta_2\right)},\dots, e^{j\left(kd_M\cos(\phi_\ell) + \Delta\theta_M\right)} \right]^T,
\end{equation}
where $d_m$ is the position of the $m$-th antenna, and $\Delta\theta_m$ is the additional phase shift incurred at the $m$-th antenna (to model the phase mismatch). Without loss of generality, we assume that $d_m$ and $\Delta\theta_m$ are fixed yet unknown random realizations, obtained from the distributions $\mathcal{N}\left((m-1)d, \sigma_d^2\right)$ and $\mathcal{N}\left(0, \sigma_p^2\right)$, respectively, where $\sigma_d$ and $\sigma_p$ model the standard deviations of the random antenna spacing and phase mismatch. Besides, we impose an additional constraint $d_1<d_2<\cdots<d_M$ to make sure the generated antenna positions physically meaningful.

\section{Problem Definition} \label{sec:Prob}

In this paper, we investigate the design of mmWave beamforming codebooks that are adaptive to the specific deployment (surrounding environment, user distribution, etc.) and the given base station hardware (array geometry, hardware imperfections, etc.). Given the system and channel models described in \sref{sec:System}, the SNR after combining for user $u$ can be written as
\begin{equation}\label{single_snr}
  \mathsf{SNR}_u = \frac{\left|{\bf w}^H{\bf h}_u\right|^2}{\left|{\bf w}\right|^2} \rho,
\end{equation}
with $\rho=\frac{P_x}{\sigma_n^2}$. Besides, we define the beamforming/combining gain of adopting $\bw$ as a transmit/receive beamformer for user $u$ as
\begin{equation}\label{single_bfgain}
  g_u = \left|{\bf w}^H{\bf h}_u\right|^2.
\end{equation}
If the combining vector $\bw$ is selected from a codebook $\boldsymbol{\mathcal{W}}$, with cardinality $\left|\boldsymbol{\mathcal{W}}\right|=N$, then, the maximum achievable SNR for use $u$ is obtained by the exhaustive search over the beam codebook as
\begin{equation}\label{best_snr}
\mathsf{SNR}^\star_u= \rho \max_{\bw \in \boldsymbol{\mathcal W}}  {\left|{\bf w}^H{\bf h}_u\right|^2},
\end{equation}
where we set $\left\|\bw\right\|^2=1$ as these combining weights are implemented using only phase shifters with constant magnitudes of $1/\sqrt{M}$, as described in \eqref{Analog}.
The objective of this paper is to design (learn) the beam codebook $\boldsymbol{\mathcal{W}}$ to maximize the SNR given by \eqref{best_snr} averaged over the set of the users that can be served by the base station. Let $\boldsymbol{\mathcal{H}}$ represent the set of channel vectors for all the users that can be served by the considered base station, the beam codebook design problem can be formulated as
\begin{align}\label{Prob-0}
 \boldsymbol{\mathcal W}_{\mathsf{opt}} = &\argmax\limits_{\boldsymbol{\mathcal W}} \hspace{2pt}  \frac{1}{|\boldsymbol{\mathcal{H}}|}\sum_{{\bf h}_u\in\boldsymbol{\mathcal{H}}} \left( \max_{{\bf w}_n\in\boldsymbol{\mathcal{W}}}\left| {\bf w}_n^H{\bf h}_u \right|^2 \right), \\
 & \subto  \hspace{2pt}  w_{mn} = \frac{1}{\sqrt{M}}e^{j\theta_{mn}}, ~ \forall m=1, ..., M, n=1,...,N, \label{cons-1} \\
 & \hspace{0.9cm} \theta_{mn}\in\boldsymbol{\Theta}, ~ \forall m=1, ..., M, n=1,...,N, \label{cons-2}
\end{align}
where $w_{mn}=\left[\bw_n\right]_m$ is the $m$-th element of the $n$-th beamforming vector in the codebook, $|\boldsymbol{\mathcal{H}}|=K$ is the total number of users, $\boldsymbol{\Theta}$ is the set that contains the $2^r$ possible phase shifts.
It is worth mentioning that the constraint in \eqref{cons-1} is imposed to uphold the adopted model where the analog beamformer can only perform phase shifts to the received signal, and the constraint in \eqref{cons-2} is to respect the quantized phase-shifters hardware constraint.

Due to the unknown array geometry as well as possible hardware impairments, the accurate channel state information is generally hard to acquire. This means that all the channels ${\bf h}_u\in\boldsymbol{\mathcal{H}}$ in the objective function are possibly unknown. Instead, the base station may only have access to the beamforming/combining gain $g_u$ (or equivalently, the Received Signal Strength Indicator (RSSI) reported by each user if a downlink setup is considered). Therefore, problem \eqref{Prob-0} is hard to solve in a general sense for the unknown parameters in the objective function as well as the non-convex constraint \eqref{cons-1} and the discrete constraint \eqref{cons-2}. Given that \textbf{this problem is essentially a search problem in a dauntingly huge yet finite and discrete space}, we consider leveraging the powerful exploration capability of deep reinforcement learning to efficiently search over the space to find the optimal or near-optimal solution.
Since the number of beams in the codebook is far less than the number of channels in $\boldsymbol{\mathcal{H}}$, then users sharing similar channels are expected to be served by the same beam that achieves the best average beamforming gain compared to the other beams in the codebook. Based on this idea, we consider solving the original problem \eqref{Prob-0} in two steps. First, we investigate the problem of learning an optimized beam pattern for a single user or a group of users that share similar channels in \sref{sec:BPL}, which we refer to as \textit{the beam pattern learning problem} and it can be formulated as
\begin{align}\label{Prob-1}
 {\bf w}_{\mathsf{opt}} = &\argmax\limits_{{\bf w}} \hspace{2pt}  \frac{1}{|\boldsymbol{\mathcal{H}}_s|}\sum_{{\bf h}_u\in\boldsymbol{\mathcal{H}}_s} \left| {\bf w}^H{\bf h}_u \right|^2, \\
 & \subto  \hspace{2pt}  w_{m} = \frac{1}{\sqrt{M}}e^{j\theta_{m}}, ~ \theta_{m}\in\boldsymbol{\Theta}, ~ \forall m=1, ..., M, \label{unit-1}
\end{align}
where $w_m$ is the $m$-th element of the beamforming vector and $\boldsymbol{\mathcal{H}}_s$ is the channel set that is supposed to contain a single channel or multiple similar channels.
Then, in \sref{sec:BCL}, we address the codebook design problem \eqref{Prob-0} by introducing a joint clustering, assignment, and beam pattern learning approach.

\section{Beam Pattern Learning} \label{sec:BPL}

\begin{figure}[t]
	\centering
	\includegraphics[width=1\columnwidth]{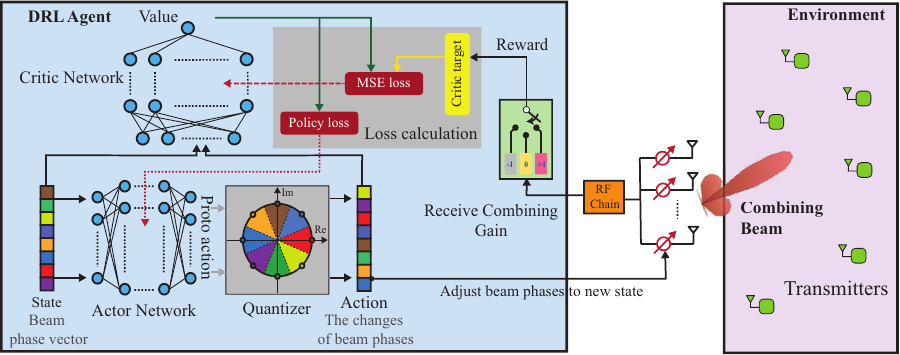}
	\caption{The proposed beam pattern design framework with deep reinforcement learning. The schematic shows the agent architecture, and the way it interacts with the environment.}
	\label{BP-Alg}
\end{figure}

In this section, we present our proposed deep reinforcement learning based algorithm for addressing the beam pattern design problem \eqref{Prob-1}, which aims at maximizing the (averaged) beamforming gain of a single user (or a group of users that share the similar channels). Given constraint \eqref{unit-1}, the design problem is essentially a search problem over a finite yet dauntingly huge discrete feasible set. For example, for a base station equipped with 32 antennas and 3-bit phase shifters, there are over $7.9\times10^{28}$ legitimate beamforming vectors. With this huge space, finding the optimal beamforming vector by using methods like exhaustive search is definitely infeasible. In order to achieve an efficient search process, we resort to reinforcement learning where the agent (base station) is able to learn from what it has experienced (receive power or feedbacks from users) and try to proceed towards a better direction (beamforming/combining vector). However, when viewing the problem from a reinforcement learning perspective, it features \textbf{finite} yet \textbf{very high dimensional} action space. This makes the traditional learning frameworks (such as deep Q-learning, deep deterministic policy gradient, etc.) hard to apply. To deal with that, we propose a learning framework based on Wolpertinger architecture \cite{Dulacarnold2015} to narrow the size of the action space and avoid missing the optimal policy at the same time.

\subsection{Wolpertinger Architecture Overview}

It is commonly known that deep Q-networks \cite{Mnih2013,Mnih2015} are very difficult to apply when the number of actions in the action space (which we refer to as the \textit{dimension} of the action space) is huge. This is because the dimension of the output of the deep Q-network relates directly to the number of possible actions, which means that the size of the neural network will keep growing as the number of actions increases.
However, for problems approaching real life complexity, it is highly likely to encounter applications that involve a huge action space, different from that in Atari games \cite{Mnih2013} where only several actions are considered.
For example, in our problem, the possible actions in the case given above (where the base station has 32 antennas and adopts 3-bit phase shifters) are in the order of $10^{28}$. The number can increase further with more antennas and higher resolution phase shifters. This is definitely intractable for the deep Q-network framework.
With this motivation, the Wolpertinger architecture is proposed as a way of reasoning in a space with large number of discrete actions \cite{Dulacarnold2015}. The Wolpertinger architecture is based on the actor-critic \cite{Sutton2018} framework and is trained using Deep Deterministic Policy Gradient (DDPG) \cite{Timothy2015}. This novel architecture utilizes a K-Nearest Neighbor (KNN) classifier to make DDPG suitable for tasks with discrete, finite yet very high dimensional action space. We briefly introduce the basic components of the Wolpertinger architecture as follows.

\subsubsection{Actor Network}

We assume an action space $\mathcal{A}\subseteq\mathbb{R}^n$ that is discrete and finite (but possibly with large number of actions), from which the agent selects an action to execute. We also assume a state space $\mathcal{S}\subseteq \mathbb{R}^m$ that contains all the possible states of an environment. We will define the action and state spaces in the context of the beam pattern learning problem in \sref{sec:DRL_learning}.
The actor network is then constructed as a function approximator parameterized by $\theta^\mu$ mapping from the state space to the $\mathbb{R}^n$, that is
\begin{equation}\label{actor}
  \mu(\cdot|\theta^\mu): \mathcal{S} \rightarrow \mathbb{R}^n.
\end{equation}
Due to the discrete and finite nature of $\mathcal{A}$, the action predicted by the actor network is probably not within $\mathcal{A}$.
In other words, for any state ${\bf s}\in\mathcal{S}$, we can get a predicted \textit{proto-action}
\begin{equation}\label{actor-pred}
  \mu({\bf s}|\theta^\mu) = \widehat{\bf a},
\end{equation}
where $\widehat{\bf a}$ is highly likely not a ``legitimate'' action, i.e. $\widehat{\bf a} \notin \mathcal{A}$. Therefore, the proto-action $\widehat{\bf a}$ needs to be transformed (quantized) to a valid action in $\mathcal{A}$, where the KNN classifier plays a role in.

\subsubsection{K-Nearest Neighbor} \label{KNN}

Since the predicted proto-action of the actor network is possibly not a valid action, we need to map $\widehat{\bf a}$ to valid actions in $\mathcal{A}$. One natural solution could be a KNN function, that is, finding $k$ actions in $\mathcal{A}$ that are most closest to $\widehat{\bf a}$ by some distance metric ($L_2$ distance to name one). More to the point, we assume that there is a function denoted by $\xi_k$. This function takes in the proto-action $\widehat{\bf a}$ and returns the $k$ nearest neighbors of that proto-action in $\mathcal{A}$ according to $L_2$ distance, formally
\begin{equation}\label{kNN}
  \xi_k(\widehat{\bf a}) = \argmin_{{\bf a}\in\mathcal{A}}^{\rm{nearest}~k} \|{\bf a} - \widehat{\bf a}\|_2.
\end{equation}
The output of $\xi_k(\widehat{\bf a})$ is a set of $k$ actions in $\mathcal{A}$ that are the top $k$ nearest neighbors to $\widehat{\bf a}$, which is denoted by $\mathcal{A}_{k} = \{{\bf a}_1, {\bf a}_2, \dots, {\bf a}_k\}$.

\subsubsection{Critic Network}

The critic network is constructed as a function approximator parameterized by $\theta^Q$ mapping from the joint state space $\mathcal{S}$ and action space $\mathcal{A}$ to $\mathbb{R}$, that is
\begin{equation}\label{critic}
  Q(\cdot,\cdot|\theta^Q): \mathcal{S}\times\mathcal{A} \rightarrow \mathbb{R}.
\end{equation}
The critic network essentially plays the role of a Q function that takes in the state and action and outputs the predicted Q value of this particular state-action pair.
Since $k$ actions are obtained from the KNN function, the critic network then evaluates $k$ state-action pairs (note that they share the same state) and selects the action that achieves the highest Q value
\begin{equation}\label{action-sel}
  {\bf a}_t = \argmax_{{\bf a}_l\in\mathcal{A}_k}Q({\bf s}_t, {\bf a}_l |\theta^Q).
\end{equation}

\subsubsection{Network Update}

The actor network aims at maximizing the output of the critic network (the predicted Q value) given a particular state, the objective of which can be simply expressed as $J(\theta^\mu) = \mathbb{E}\left[Q({\bf s},{\bf a}|_{{\bf a}=\mu({\bf s}|\theta^\mu)})\right]$.
Thus, the actor policy is updated using the deep deterministic policy gradient, which is given by
\begin{align}\label{actor-policy}
  -\nabla_{\theta^\mu}J(\theta^\mu) \approx & -\mathbb{E}\left[\nabla_{{\bf a}}Q({\bf s}, {\bf a})\nabla_{\theta^\mu}\mu({\bf s}|\theta^\mu)\right] \\
  \approx & -\frac{1}{B}\sum_{b=1}^{B} \nabla_{{\bf a}}Q({\bf s}, {\bf a})|_{{\bf s}={\bf s}_b, {\bf a}=\mu({\bf s}_b|\theta^\mu)}\nabla_{\theta^\mu}\mu({\bf s}|\theta^\mu)|_{{\bf s}={\bf s}_b}. \label{updateStep}
\end{align}
The objective of the critic network is to estimate the Q value of the input state-action pair. Thus, the target can be constructed in the exactly same way that is adopted in the deep Q-networks, which is given by
\begin{equation}\label{critic-target}
  y = \mathbb{E}\left[ r + \gamma \max_{{\bf a}_{t+1}}Q({\bf s}_{t+1}, {\bf a}_{t+1} |\theta^Q) \right].
\end{equation}
The parameters of the critic network $\theta^Q$ is then updated based on the mean squared error over a particular mini-batch, which is given by $\frac{1}{B}\sum_{b=1}^{B}\left(y_b-Q({\bf s}_b, {\bf a}_b |\theta^Q)\right)^2$.

For the sake of computational stability, the actor and critic networks have duplicates, referred to as the target actor and target critic networks. They are not trainable like the actor and critic networks, but they are utilized for calculating the targets. Despite them being not trainable, the parameters of the target actor and critic networks get updated using the parameters of the critic and actor networks after a certain number of training iterations. Formally, it can be expressed as
\begin{align}
  \theta^{Q^\prime} \leftarrow & \tau\theta^Q + (1-\tau)\theta^{Q^\prime}, \label{t1} \\
  \theta^{\mu^\prime} \leftarrow & \tau\theta^\mu + (1-\tau)\theta^{\mu^\prime}, \label{t2}
\end{align}
where $\theta^{\mu^\prime}$ and $\theta^{Q^\prime}$ are the parameters of target actor network and target critic network, $\tau$ is a non-negative hyper-parameter usually taking a value far less than $1$.

\subsection{DRL Based Beam Pattern Design}\label{sec:DRL_learning}

In this subsection, we describe in detail our proposed DRL based beam pattern design approach. We adopt the Wolpertinger architecture described above as our learning framework.

\subsubsection{Reinforcement Learning Setup}

To solve the problem with reinforcement learning, we first specify the corresponding building blocks of the learning algorithm.
\begin{itemize}
  \item \textbf{State:} We define the state ${\bf s}_t$ as a vector that consists of the phases of all the phase shifters at the $t$-th iteration, that is, ${\bf s}_t=\left[\theta_1, \theta_2, \dots, \theta_M\right]^T$. This phase vector can be converted to the actual beamforming vector by applying \eqref{Analog}. Since all the phases in ${\bf s}_t$ are selected from $\boldsymbol\Theta$, and all the phase values in $\boldsymbol\Theta$ are within $(-\pi, \pi]$, \eqref{Analog} essentially defines a bijective mapping from the phase vector to the beamforming vector. Therefore, for simplicity, we will use the term ``beamforming vector'' to refer to both this phase vector and the actual beamforming vector (the conversion is by \eqref{Analog}), according to the context.
  \item \textbf{Action:} We define the action ${\bf a}_t$ as the element-wise changes to all the phases in ${\bf s}_t$. Since the phases can only take values in $\boldsymbol\Theta$, a change of a phase means that the phase shifter selects a value from $\boldsymbol\Theta$. Therefore, the action is directly specified as the next state, i.e. ${\bf s}_{t+1}={\bf a}_t$. 
  \item \textbf{Reward:} We define a ternary reward mechanism, i.e. the reward $r_t$ takes values from $\{+1, 0, -1\}$. We compare the beamforming gain achieved by the current beamforming vector, denoted by $g_t$, with two values: (i) an adaptive threshold $\beta_t$, and (ii) the previous beamforming gain $g_{t-1}$. The reward is computed using the following rule
      \begin{itemize}
        \item $g_t > \beta_t$, $r_t=+1$;
        \item $g_t \le \beta_t$ and $g_t > g_{t-1}$, $r_t=0$;
        \item $g_t \le \beta_t$ and $g_t \le g_{t-1}$, $r_t=-1$.
      \end{itemize}
\end{itemize}

We adopt an adaptive threshold mechanism that does not rely on any prior knowledge of the channel distribution. The threshold has an initial value of zero. When the base station tries a beam and the resulting beamforming/combining gain surpasses the current threshold, the system updates the threshold by the value of this beamforming/combining gain. Besides, because the update of threshold also marks a successful detection of a new beam that achieves the best beamforming/combining gain so far, the base station also records this beamforming vector. As can be seen in this process, in order to evaluate the quality of a beam (or equivalently, calculate the reward), the system always tracks two quantities, which are the previous beamforming/combining gain and the best beamforming/combining gain achieved so far (i.e. the threshold).

\begin{algorithm}[ht]
	\caption{DRL Based Beam Pattern Learning}
	\label{alg1}
	\begin{algorithmic}[1]
    \STATE Initialize actor network $\mu({\bf s}|\theta^\mu)$ and critic network $Q({\bf s}, {\bf a}|\theta^Q)$ with random weights $\theta^\mu$ and $\theta^Q$
    \STATE Initialize target networks $\mu^\prime$ and $Q^\prime$ with the weights of actor and critic networks' $\theta^{\mu^\prime}\leftarrow\theta^\mu$ and $\theta^{Q^\prime}\leftarrow\theta^Q$
    \STATE Initialize the replay memory $\mathcal{D}$, minibatch size $B$, discount factor $\gamma$
    \STATE Initialize adaptive threshold $\beta=0$ and the previous average beamforming gain $g_1=0$
    \STATE Initialize a random process $\mathcal{N}$ for action exploration
    \STATE Initialize a random beamforming vector ${\bf w}_1$ as the initial state ${\bf s}_1$
    \FOR{$t=1$ to $T$}
        \STATE Receive a proto-action from actor network with exploration noise $\widehat{\bf a}_t = \mu({\bf s}_t|\theta^\mu) + \mathcal{N}_t$
        \STATE Quantize the proto-action to a valid beamforming vector ${\bf a}_t$ according to \eqref{quant}
        \STATE Execute action ${\bf a}_t$, observe reward $r_t$ and update state to ${\bf s}_{t+1}={\bf a}_t$
        \STATE Update the threshold $\beta$ and the previous beamforming gain $g_t$
        \STATE Store the transition $({\bf s}_t, {\bf a}_t, r_t, {\bf s}_{t+1})$ in $\mathcal{D}$
        \STATE Sample a random mini batch of $B$ transitions $({\bf s}_b, {\bf a}_b, r_b, {\bf s}_{b+1})$ from $\mathcal{D}$
        \STATE Calculate target $y_b=r_b+\gamma Q^\prime({\bf s}_{b+1}, \mu^\prime({\bf s}_{b+1}|\theta^{\mu^\prime})|\theta^{Q^\prime})$
        \STATE Update the critic network by minimizing the loss $L=\frac{1}{B}\sum_{b}(y_b-Q({\bf s}_b, {\bf a}_b|\theta^Q))^2$
        \STATE Update the actor network using the sampled policy gradient given by \eqref{updateStep}
        \STATE Update the target networks every $C$ iterations by \eqref{t1} and \eqref{t2}
    \ENDFOR
	\end{algorithmic}
\end{algorithm}

\subsubsection{Environment Interaction} \label{LF}

As mentioned in Sections \ref{intro} and \ref{sec:Prob}, due to the possible hardware impairments, accurate channel state information is generally unavailable. Therefore, the base station can only resort to the receive power (or beamforming gain feedback reported by the users in a downlink setup) to adjust its beam pattern in order to achieve a better performance. To be more specific, upon forming a new beam $\tilde{\bw}$, the base station uses this beam to receive the symbols transmitted by every user. Then, it averages all the combining gains as follows
\begin{equation}\label{avg_bf_eval}
  \bar{g} = \frac{1}{|\boldsymbol{\mathcal{H}}_s|}\sum_{{\bf h}_u\in\boldsymbol{\mathcal{H}}_s} \left| \tilde{\bw}^H{\bf h}_u \right|^2,
\end{equation}
where $\boldsymbol{\mathcal{H}}_s$ represents the targeted user channel set. Recall that \eqref{avg_bf_eval} is the same as evaluating the objective function of \eqref{Prob-1} with the current beamforming vector $\tilde{\bw}$. Depending on whether or not the new average beamforming/combining gain surpasses the previous beamforming/combining gain as well as the current threshold, the base station gets either reward or penalty, based on which it can judge the ``quality'' of the current beam and decide how to move.

\subsubsection{Exploration} \label{subsub:Explore}

The exploration happens after the actor network predicts the proto-action $\widehat{\bf a}_{t+1}$ based on the current state (beam) ${\bf s}_{t}$. Upon obtaining the proto-action, an additive noise is added element-wisely to $\widehat{\bf a}_{t+1}$ for the purpose of exploration, which is a customary way in the context of reinforcement learning with continuous action spaces \cite{Sutton2018,Timothy2015}. In our problem, we use temporally correlated noise samples generated by an Ornstein-Uhlenbeck process \cite{Uhlenbeck1930}, which is also used in \cite{Timothy2015,Dulacarnold2015}. It is worth mentioning that a proper configuration of the noise generation parameters has significant impact on the learning process. Normally, the extent of exploration (noise power) is set to be a decreasing function with respect to the iteration number, which is commonly known as exploration-exploitation tradeoff \cite{Sutton2018}.
Furthermore, the exact configuration of noise power should relate to specific applications. In our problem, for example, the noise is directly added to the predicted phases. Thus, at the very beginning, the noise should be strong enough to perturb the predicted phase to any other phases in $\boldsymbol\Theta$. By contrast, when the learning process approaches to the termination (the learned beam already performs well), the noise power should be decreased to a smaller level that is only capable of perturbing the predicted phase to its adjacent phases in $\boldsymbol\Theta$.

\subsubsection{Quantization} \label{subsub:Quantize}

The ``proto'' beam (with exploration noise added) should be quantized in order to be a valid new beam. To this end, we apply a KNN classifier as described in \sref{KNN}. Furthermore, we specify $k=1$ in \eqref{kNN}, which is basically a nearest neighbor lookup. Therefore, each quantized phase in the new vector can be simply calculated as
\begin{equation}\label{quant}
  [{\bf s}_{t+1}]_m = \argmin_{\theta\in\Theta}\left|\theta-[\widehat{{\bf s}}_{t+1}]_m\right|, \forall m=1, 2, \dots, M.
\end{equation}

\subsubsection{Forward Computation and Backward Update} \label{subsub:FB}

The current state ${\bf s}_{t}$ and the new state ${\bf s}_{t+1}$ (recall that we directly set ${\bf s}_{t+1}={\bf a}_t$) are then fed into the critic network to compute the Q value, based on which the targets of both actor and critic networks are calculated. This completes a forward pass. Following that, a backward update is performed to the parameters of the actor and critic networks. A pseudo code of the algorithm can be found in \aref{alg1}.

\section{Beam Codebook Learning} \label{sec:BCL}

In this section, we propose a multi-network DRL approach for solving \eqref{Prob-0} and learning a beam codebook. The solution is built around the beam pattern learning approach described in \sref{sec:BPL}. It could be briefly described as a pipeline of three key stages, namely clustering, assignment, and beam pattern learning. The first stage learns to \textit{partition} the users in the environment into clusters based on how similar their channels are (without explicitly estimating those channels). These clusters are, then, \textit{assigned} to different DRL networks in the second stage. Such assignment needs to guarantee a form of consistency among the clusters that are assigned to the networks during the learning process. Finally, the third stage is where \textit{the beam pattern learning} happens. Each of the DRL networks is expected to learn a beam pattern, and the collection of those patterns constructs the beam codebook. We detail this approach in the following three subsections.

\begin{figure}[t]
	\centering
	\includegraphics[width=1\columnwidth]{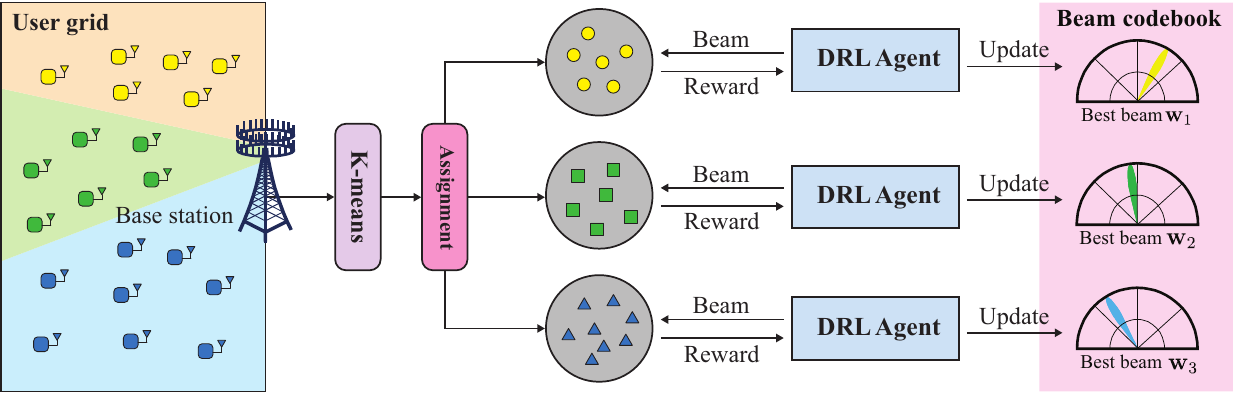}
	\caption{The proposed beam codebook design framework with deep reinforcement learning. It shows the overall architecture, including the user clustering component.}
	\label{BC-Alg}.
\end{figure}

\subsection{User Clustering} \label{U-cluster}

Following into the footsteps of \cite{Zhang2020}, users sharing similar channels are served by the same beam in the codebook. The question then becomes how to cluster the users' channels without knowing them, i.e., without performing expensive channel estimation. As a result of the constant modulus and limited resolution phase shifters, the set of feasible beamforming vectors for \eqref{Prob-0} forms a huge yet finite subset of $\mathbb{C}^M$, and all those vectors live on the surface of the $M$-dimensional unit hypersphere. The proposed clustering method here relies on utilizing a random subset of those vectors, henceforth referred to as \textit{the sensing beams}, for the purpose of gathering sensing information in the form of receive combining gain. This information is used to cluster those users, developing a rough sense of their distribution in the environment.

To perform the clustering, the method starts by constructing a matrix that is comprised of receive combining gains using the sensing beams. Formally, let $\mathcal{F}=\{{\bf f}_1, {\bf f}_2, \dots, {\bf f}_S\}$ be a set of $S$ sensing beams that are randomly sampled from the feasible set of \eqref{Prob-0}\footnote{To clarify, we use $\boldsymbol{\mathcal{W}}$ and ${\bf w}$ to denote the learned codebook and beam. We use $\mathcal{F}$ and ${\bf f}$ to denote the sensing beam set and sensing beam.}, where $\mathbf f_s\in \mathbb C^M, \ \forall s\in \{1,\dots,S\}$. Also, let $\boldsymbol{\mathcal{H}}_{\mathrm{sen}}=\{{\bf h}_1, {\bf h}_2, \dots, {\bf h}_{K^\prime}\}$ denote the channels of the $K^\prime$ users that contribute to the clustering process, where $\boldsymbol{\mathcal{H}}_{\mathrm{sens}} \subseteq \boldsymbol{\mathcal{H}}$.
It is worth mentioning that these $K^\prime$ users do not need to be present in the environment at the same time. The receive combining gains used in the clustering algorithm can be collected over a relatively long period of time. This is because essentially the learned clustering is a function of the major elements (e.g. walls, buildings, large trees, etc.) of the environment. Such scatterers/reflectors commonly stay static over long periods of time. As a result, the sensing data can be collected in an accumulative manner and the learned classifier does not need to be updated (re-trained) frequently.
The objective then is to collect receive combining gains form the $K^\prime$ users for every beam $\bf f_s \in \mathcal F$. More specifically, the receive combining gains are used to construct the \emph{sensing matrix} ${\bf P}$
\begin{equation}\label{bfMatrix}
  {\bf P}\triangleq\left[\begin{array}{ccccc}
          \left|{\bf f}_1^H{\bf h}_1\right|^2 & \cdots & \left|{\bf f}_1^H{\bf h}_k\right|^2 & \cdots & \left|{\bf f}_1^H{\bf h}_{K^\prime}\right|^2 \\
          \left|{\bf f}_2^H{\bf h}_1\right|^2 & \cdots & \left|{\bf f}_2^H{\bf h}_k\right|^2 & \cdots & \left|{\bf f}_2^H{\bf h}_{K^\prime}\right|^2 \\
          \vdots & \ddots & \vdots & \ddots & \vdots \\
          \left|{\bf f}_S^H{\bf h}_1\right|^2 & \cdots & \left|{\bf f}_S^H{\bf h}_k\right|^2 & \cdots & \left|{\bf f}_S^H{\bf h}_{K^\prime}\right|^2
        \end{array}\right],
\end{equation}
where each column in ${\bf P}$ has the receive combining gains from the same user for all sensing beams in $\mathcal{F}$. It is worth mentioning that since the receive combining gain is the \textbf{only} information source to the base station, the sensing matrix ${\bf P}$ actually incorporates \textbf{all} the information that the base station can leverage from the outside environment.

The sensing matrix is used to extract feature vectors that characterize the user distribution in the environment. Each column in $\mathbf P$ represents the receive gains of a single user in the environment. One could cluster the users by directly applying a clustering algorithm (such as k-means) on the columns of $\mathbf P$. However, empirical evidence shows that this clustering does not yield meaningful partitioning of the users (or equivalently the channels). The reason for that could be attributed to the fact that the columns of $\mathbf P$ are restricted to the nonnegative orthant of the $\mathbb R^S$ vector space; this increases the likelihood of overlapping clusters, which are hard to separate with k-means. As an alternative, we propose to transform the column of $\mathbf P$ using pair-wise differences. More precisely, the pair-wise differences of the elements of every column are computed, scaled, and stacked in a column vector as follows
\begin{equation}\label{fvec}
  {\bf u}_k = \left(\frac{1}{S}\sum_{s=1}^{S}\left|{\bf f}_s^H{\bf h}_k\right|^2\right)^{-1} \left[\begin{array}{c}
                             \left|{\bf f}_1^H{\bf h}_k\right|^2-\left|{\bf f}_2^H{\bf h}_k\right|^2 \\
                             \left|{\bf f}_1^H{\bf h}_k\right|^2-\left|{\bf f}_3^H{\bf h}_k\right|^2 \\
                             \vdots \\
                             \left|{\bf f}_{S-1}^H{\bf h}_k\right|^2-\left|{\bf f}_S^H{\bf h}_k\right|^2
                           \end{array}\right]_{\frac{S(S-1)}{2}\times 1}, ~ \forall k=1,2,\dots,K^\prime,
\end{equation}
where ${\bf u}_k$ is referred to as the \emph{feature vector} of user $k$. The column vectors of all $K^{\prime}$ users are organized in a feature matrix $\bf U=[{\bf u}_1, {\bf u}_2, \dots, {\bf u}_{K^\prime}]$. This choice of transformation preserves the relation between the channel vector of a user and the sensing vectors, i.e., the sense of how close a channel vector to each sensing vector. However, it expresses that relation using a feature vector that could fall anywhere in the $\mathbb R^{S(S-1)/2}$ vector space (not restricted to the nonnegative orthant). The factor in \eqref{fvec} expresses each elements in the columns of $\bf U$ as a ratio of a pair-wise difference to the average power of the corresponding column of matrix $\mathbf P$.

The clustering is applied on the columns of the feature matrix ${\bf U}$ to produce $N$ clusters. The popular k-means algorithm \cite{PatternRecog} is adopted to generate those clusters. The algorithm learns to partition the $K^\prime$ users (or equivalently their channels $\mathcal H_{\mathrm{sen}}$) into $N$ disjoint subsets
\begin{equation}\label{clustering}
  \boldsymbol{\mathcal{H}}_{\mathrm{sen}} = \boldsymbol{\mathcal{H}}_1 \cup \boldsymbol{\mathcal{H}}_2 \cup \cdots \cup \boldsymbol{\mathcal{H}}_N,
\end{equation}
where $\boldsymbol{\mathcal{H}}_k\cap\boldsymbol{\mathcal{H}}_l=\emptyset, \forall k\ne l$ and we assume that the subscript of each user group is also the corresponding label of that group. The trained k-means algorithm is used to classify any new user coming into the environment. It is important to note here that the learned clustering is a function of the major elements of the environment not the user distribution, i.e., it is mainly affected by major scatterers and their positions like walls, buildings, large trees, etc. Such scatterers commonly change over long periods of time, and consequently, the learned clusters do not need to be updated frequently.

\subsection{Cluster Assignment} \label{G-assign}

Since the clustering will be frequently repeated whenever there is a change in the environment, an important question arises: how to assign the new clusters to the existing DRL networks, with each of them learning one beam? The answer to this question defines the second stage in our proposed codebook learning approach. For the learning process to be meaningful, a network should consistently be assigned channel clusters that exhibit some form of similarity; the new cluster should be similar to the previous one in the sense that the network can improve its currently learned beam pattern but not change it completely. To that end, we formulate this cluster assignment task as a linear sum assignment problem, which can be solved efficiently using the \emph{Hungarian algorithm} \cite{HungarianAlgo}. In such problem, every pair of new cluster and DRL network is assigned a cost reflecting how suitable this cluster to the network, and the goal is to find \textbf{$N$ unique cluster-network assignments} that minimize the total cost sum (total suitability).

To perform the cluster-network assignment, a cost needs to be computed to measure suitability and guide the assignment process. Let $\widehat{\boldsymbol{\mathcal{H}}}_{\mathrm{sen}} = \widehat{\boldsymbol{\mathcal{H}}}_1 \cup \widehat{\boldsymbol{\mathcal{H}}}_2 \cup \cdots \cup \widehat{\boldsymbol{\mathcal{H}}}_N$ be the new clusters obtained using the clustering algorithm described in \sref{U-cluster}. As described in \sref{LF}, the DRL network always tracks the beamforming vectors that achieve the best beamforming gain, which forms a set of ``temporarily best'' beamforming vectors, denoted by $\Upsilon = \{\widehat{\bf w}_1, \widehat{\bf w}_2, \dots, \widehat{\bf w}_N\}$, where the subscripts stand for the indices of the $N$ DRL networks. We propose to use the average beamforming gain of each beamforming vector in $\Upsilon$ computed on each cluster as the suitability measure. The result of that forms a cost matrix $\mathbf{Z}$, where the value at the intersection of $n$-th row and $n^\prime$-th column of $\mathbf{Z}$ stands for the average beamforming gain of the $n$-th temporarily best beamforming vector in $\Upsilon$ on the $n^\prime$-th channel cluster in $\widehat{\boldsymbol{\mathcal{H}}}_{\mathrm{sen}}$. This value is calculated as
\begin{equation}\label{inter_gain}
  \mathbf{Z}_{nn^\prime} = \frac{1}{|\widehat{\boldsymbol{\mathcal{H}}}_{n^\prime}|}\sum_{{\bf h}\in\widehat{\boldsymbol{\mathcal{H}}}_{n^\prime}}\left| \widehat{\bf w}_n^H {\bf h} \right|^2.
\end{equation}

With the cost matrix, we formulate the cluster assignment task as a linear sum assignment problem, which is given by
\begin{align}\label{LSAP}
  \min_{\mathbf{X}} & ~ -\sum_{n=1}^{N}\sum_{n^\prime=1}^{N}\mathbf{X}_{nn^\prime}\mathbf{Z}_{nn^\prime} \\
  \subto & ~ \mathbf{X} ~ \text{is a permutation matrix}.
\end{align}
This problem can be efficiently solved by using Hungarian algorithm, the results of which are $N$ association tuples
\begin{equation}\label{repBeam}
  (\widehat{\bf w}_n, \widehat{\boldsymbol{\mathcal{H}}}_{n^\prime}), ~ n, n^\prime\in\{1,2,\dots,N\}. \notag
\end{equation}
In other words, the cluster assignment step forms a bijective mapping from $\Upsilon$ to the set of channel groups
\begin{equation}\label{bijective}
  \{\widehat{\bf w}_1, \widehat{\bf w}_2, \dots, \widehat{\bf w}_N\} \Longleftrightarrow
  \{\widehat{\boldsymbol{\mathcal{H}}}_1, \widehat{\boldsymbol{\mathcal{H}}}_2, \dots, \widehat{\boldsymbol{\mathcal{H}}}_N\}.
\end{equation}

\begin{algorithm}[ht]
	\caption{User Clustering and Cluster Assignment Algorithm}
	\label{alg2}
	\begin{algorithmic}[1]
    \STATE Initialize a sensing beam set $\mathcal{F}=\{{\bf f}_1, {\bf f}_2, \dots, {\bf f}_S\}$
    \STATE Initialize the temporarily best beam set $\Upsilon = \{\widehat{\bf w}_1, \widehat{\bf w}_2, \dots, \widehat{\bf w}_N\}$
    \STATE Construct sensing matrix ${\bf P}$ by \eqref{bfMatrix}
    \STATE Transform sensing matrix ${\bf P}$ to feature matrix ${\bf U}$ by applying \eqref{fvec} to the columns of ${\bf P}$
    \STATE Use k-means algorithm to cluster the columns of ${\bf U}$ into $N$ clusters
    \WHILE{environment has not changed}
        \STATE Randomly sample a subset of users $\widehat{\boldsymbol{\mathcal{H}}}$ from $\boldsymbol{\mathcal{H}}$
        \STATE Partition sampled channels using the trained k-means classifier to $\widehat{\boldsymbol{\mathcal{H}}} = \widehat{\boldsymbol{\mathcal{H}}}_1 \cup \widehat{\boldsymbol{\mathcal{H}}}_2 \cup \cdots \cup \widehat{\boldsymbol{\mathcal{H}}}_N$ channel clusters
        \STATE Construct the matrix $\mathbf{Z}$ using $\Upsilon$ and the clustering result of $\widehat{\boldsymbol{\mathcal{H}}}$ based on \eqref{inter_gain}
        \STATE Solve the optimization problem \eqref{LSAP} by applying Hungarian algorithm
        \STATE Assign the user clusters to DRL networks based on the association relationship given by the permutation matrix $\mathbf{X}$
        \STATE Train the $N$ DRL networks
        \IF{training saturated}
        \STATE Fine-tune the learned beam pattern using perturb-and-quantize operations
        \ENDIF
    \ENDWHILE
    \STATE Go to line 1
	\end{algorithmic}
\end{algorithm}

\subsection{Neural Network Update and Fine-Tuning}

Upon obtaining the clustered channels and their assignment \eqref{bijective}, the problem \eqref{Prob-0} is essentially decomposed into $N$ independent sub-problems which is given by \eqref{Prob-1}. Each DRL network adjusts its own beam based on the assigned user cluster. They only consider the receive combining gains from their designated users. User clustering and cluster assignment are two key stages that enable adaptability and empower the proposed solution with capability of dealing with dynamic environment. Practically speaking, it is impossible to fix all the users until a good beam codebook is learned. Instead, we keep learning cluster and assign the users as they change over time, which partially reflects the dynamics of the environment. Our proposed beam codebook approach accounts for such practical considerations and is able to learn beam codebooks that adapt to the environment. The complete beam codebook learning algorithm is given in \aref{alg2}.

The beam pattern learning proceeds as described in \sref{sec:DRL_learning} with one minor difference, a final perturb-and-quantize fine-tuning step. This step is basically applied after the DRL agent reaches training saturation. It is composed of three simple operations: (i) perturb the beam vector with exploration noise, (ii) quantize the perturb beam vector, and (iii) evaluate the quantized beam vector on the assigned cluster of users. The training algorithm loops over the three operations until the received beamforming gain saturates again. The goal of this last stage is to fine-tune the beam pattern without the relatively expensive agent-training process.

\section{Experiments Setup and Network Training} \label{sec:Exps}

To evaluate the performance of the proposed solutions, two scenarios are considered. They are designed to represent two different communication settings. The first has all users experiencing LOS connection with the BS, while the other has them experiencing NLOS connection. The following two subsections provide more details on the scenarios and the training process.

\begin{figure}[t]
	\centering
	\subfigure[LOS Scenario]{ \includegraphics[width=0.45\linewidth]{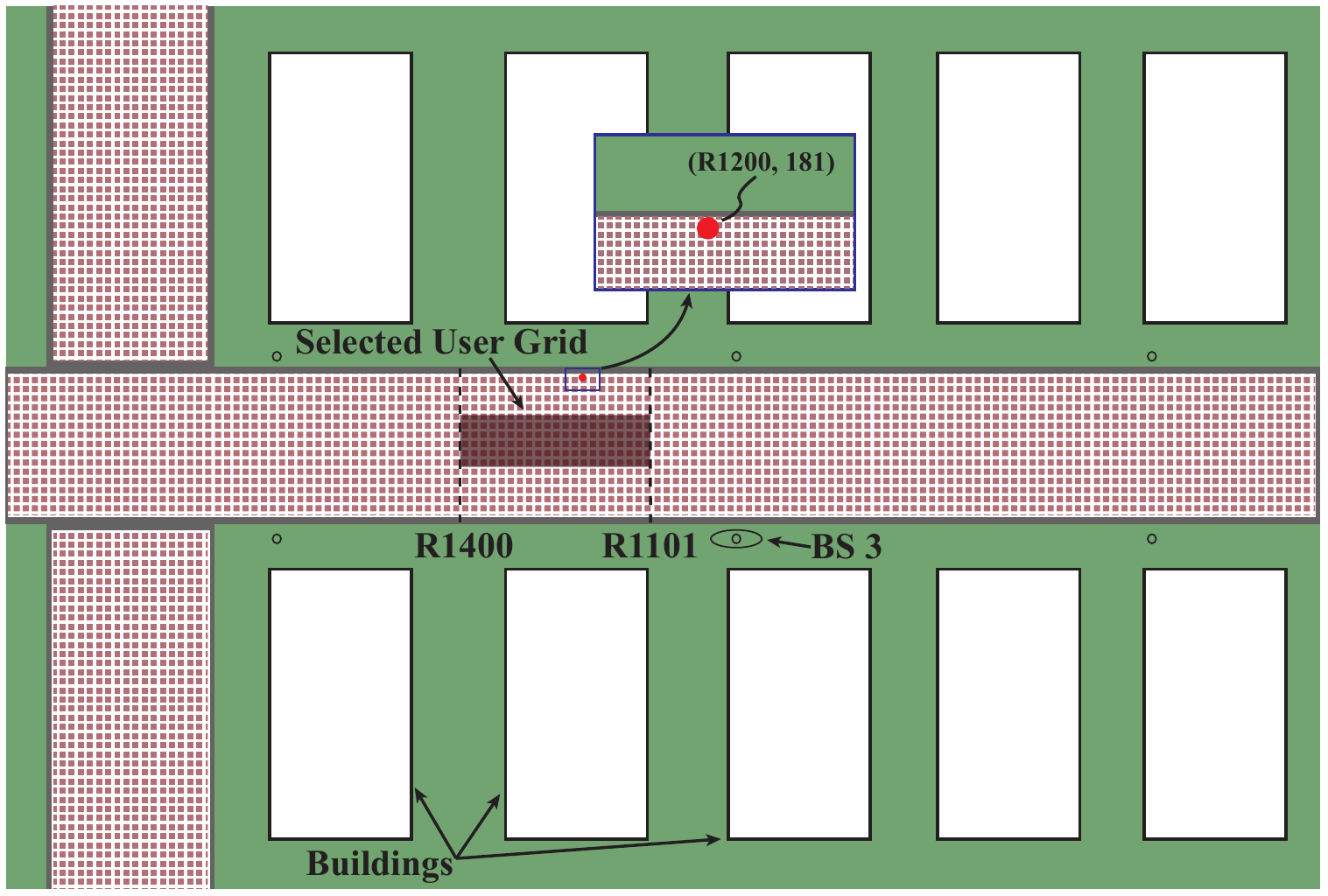}\label{fig:sce_los} }
	\subfigure[NLOS Scenario]{ \includegraphics[width=0.45\linewidth]{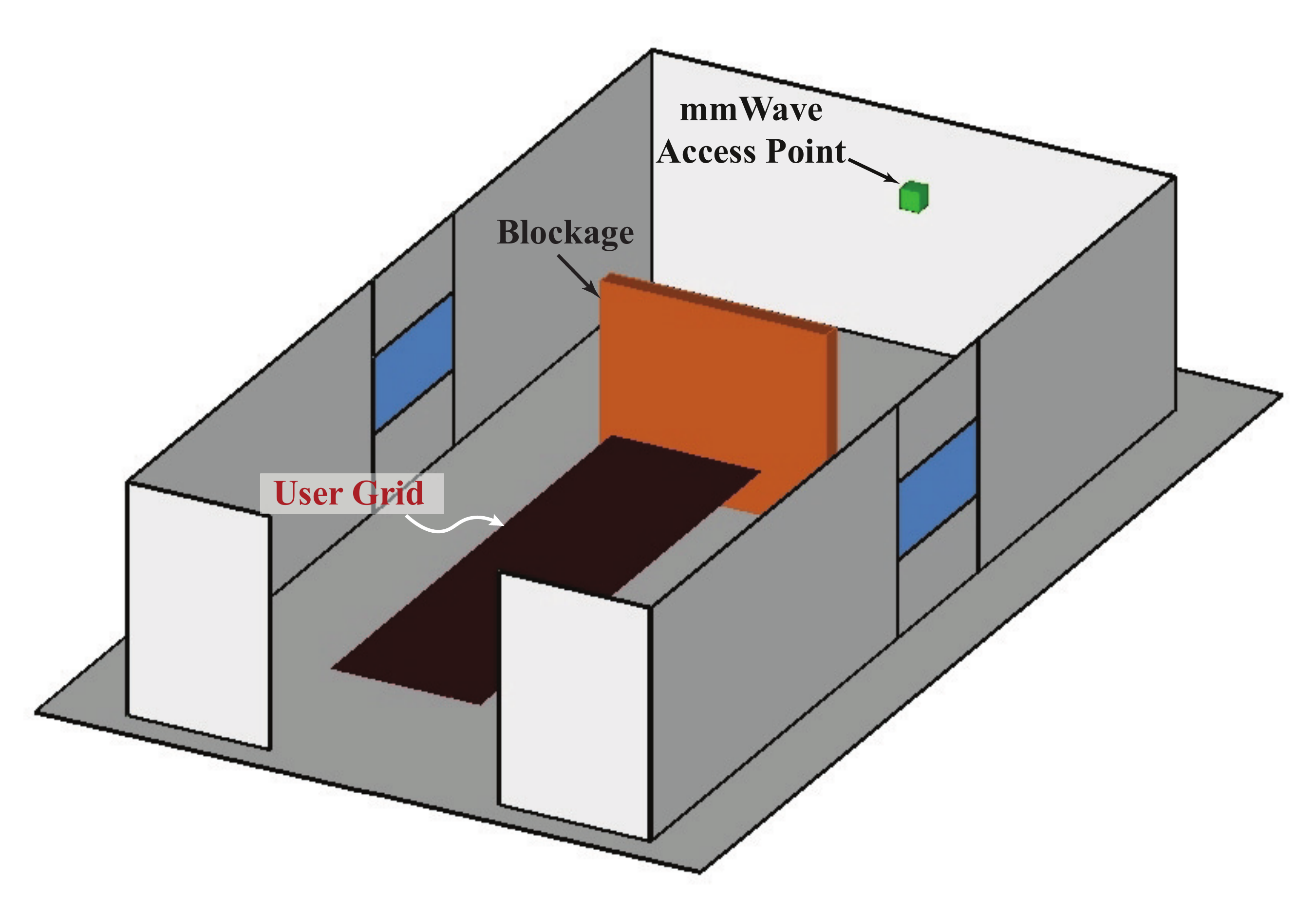}\label{fig:sce_nlos} }
	\caption{Two perspective views of the considered communication scenarios. (a) shows the LOS scenario. It is chosen to be outdoor since the likelihood of LOS connection is higher there. (b) shows the NLOS scenario. Similar to (a), this scenario has been chosen for the high likelihood of having NLOS users indoors.}
	\label{scenarios}
\end{figure}

\subsection{Communication Scenarios and Datasets}

Two scenarios are used for performance evaluation, as shown in \fref{scenarios}. The first one is an outdoor LOS scenario where all users have LOS connection with the mmWave BS, with an operating frequency of 60 GHz. The second one is chosen to be an indoor NLOS scenario where all the users have NLOS connection with the mmWave BS, with an operating frequency of 28 GHz. Both scenarios are part of DeepMIMO dataset \cite{DeepMIMO}. Using the DeepMIMO scripts, two sets of channels, namely $\mathcal{S}^{\text{LOS}}$ and $\mathcal{S}^{\text{NLOS}}$, are generated, one for each scenario. Table \ref{param} shows the data generation hyper-parameters.
The datasets taking into account the hardware impairments are generated based on the LOS scenario. While our proposed solution can deal with general impairments, we only consider two main sources of impairments, namely antenna spacing and phase mismatches. We generate multiple datasets based on different levels of impairments, measured by the standard deviations of antenna spacing and phase mismatches. Without distinction of them, we denote those datasets with impairments as $\mathcal{S}^{\text{cLOS}}$\footnote{``cLOS'' is a shorthand of corrupted LOS.}.

\subsection{Machine Learning Model Structure and Pre-processing}

While we generate multiple datasets, the learning architecture is the same, which is based on the DDPG framework. It is comprised of two networks, actor and critic. The input of the actor network is the state, i.e. the phases of the phase shifters, hence with a dimension of $M$. There are two hidden layers, all comprising $16M$ neurons and followed by Rectified Linear Unit (ReLU) activations. The output of the actor network is the predicted action, which also has a dimension of $M$ and is followed by hyperbolic tangent (tanh) activations scaled by $\pi$. For the critic network, the input is the concatenation of the state and action, so it has a dimension of $2M$. There are also two hidden layers, all with $32M$ neurons and followed by ReLU activations. The output of the critic network stands for the predicted Q value of the input state-action pair, which is a real scalar (dimension of 1). The hyper-parameters for training can be found in Table \ref{train_hyp}.
The training process starts by data pre-processing. The channels in each dataset are normalized to improve the training experience, which is a very common practice in machine learning \cite{EffBackProp}. As in \cite{Zhang2020,Zhang2020Deep}, the channel normalization using the maximum absolute value in the dataset helps the network undergo a stable and efficient training. Formally, the normalization factor is found as follows
\begin{equation}\label{normalization}
  \Delta = \max_{{\bf h}_u\in\mathcal{S}}\left|[{\bf h}_u]_m\right|,
\end{equation}
where $\mathcal{S}\in\{\mathcal{S}^{\text{LOS}}, \mathcal{S}^{\text{NLOS}}, \mathcal{S}^{\text{cLOS}}\}$ and $[{\bf h}_u]_m$ is the $m$-th element in the channel vector ${\bf h}_u$.

\begin{table}[t]
	\begin{minipage}{.55\linewidth}
		\caption{Hyper-parameters for channel generation}
		\centering
		\begin{tabular}{|c | c | c|}
			\hline
			Parameter & \multicolumn{2}{c|}{value} \\
			\hline\hline
			Name of scenario & O1\textunderscore60 & I2\textunderscore28B \\
			\hline
			Active BS & 3 & 1 \\
			\hline
			Active users & 1101 to 1400 & 201 to 300 \\
			\hline
			Number of antennas (x, y, z)  & (1, 32, 1) & (32, 1, 1) \\
			\hline
			System BW & 0.5 GHz & 0.5 GHz\\
			\hline
			Antenna spacing & 0.5 & 0.5 \\
			\hline
			Number of OFDM sub-carriers & 1 & 1 \\
			\hline
			OFDM sampling factor & 1 & 1 \\
			\hline
			OFDM limit & 1 & 1 \\
			\hline
			Number of multi paths & 5 & 5 \\
			\hline
		\end{tabular}
		\label{param}
	\end{minipage}
	\hfil
	\begin{minipage}{.45\linewidth}
		\caption{Hyper-parameters for model training}
		\centering
		\begin{tabular}{|c | c | c|}
			\hline
			Parameter & \multicolumn{2}{c|}{value} \\
			\hline\hline
			Models & Actor & Critic \\
			\hline
			Replay memory size & 8192 & 8192 \\
			\hline
			Mini-batch size & 1024 & 1024 \\
			\hline
			Learning rate & $10^{-3}$ & $10^{-3}$ \\
			\hline
			Weight decay & $10^{-2}$ & $10^{-3}$ \\
			\hline
		\end{tabular}
		\label{train_hyp}
	\end{minipage}
\end{table}

\section{Simulation Results} \label{sec:Results}

In this section, we evaluate the performance of the proposed solution using the scenarios described in \sref{sec:Exps}. In a nutshell, the numerical results show that our proposed learning solutions can adapt to different environments, user distributions as well as hardware impairments, \textbf{without the need to estimate the channels.} We compare the performance of the learned codebook with classical beamsteering codebook, where the beamforming vectors are spatial matched filters for the single-path channels. Therefore, they have the same form of the array response vector and can be parameterized by a simple angle \cite{Alkhateeb2015Limited}. In our simulation, depending on the adopted size of the classical beamsteering codebook, those angles are evenly spaced in the range of $[0, \pi]$. Next, we will first evaluate the performance of the beam pattern learning solution in \sref{subsec:simu-BPL}, and then evaluate the beam codebook learning solution in \sref{subsec:simu-BCL}.

\begin{figure}[t]
	\centering
	\includegraphics[width=.95\columnwidth]{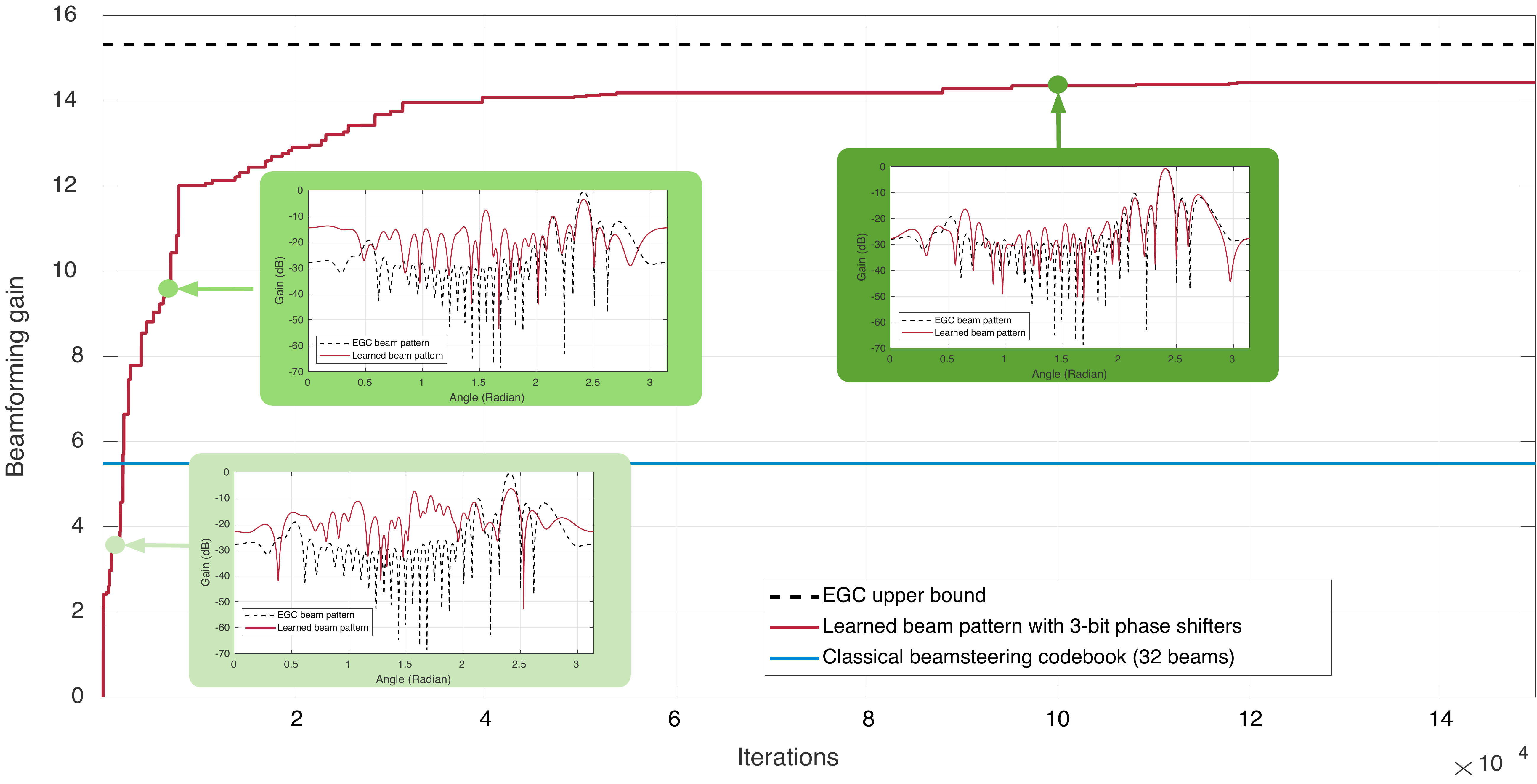}
	\caption{The beam pattern learning results for a single user with LOS connection to the base station. The base station employs a perfect uniform linear array with 32 antennas and 3-bit phase shifters. In this figure, we show the learning process and the beam patterns learned at three different stages during the iterations. The learned beam patterns are plotted using solid red line, and the equal gain combining/beamforming vector is plotted using dashed black line.}
	\label{1b-clean}.
\end{figure}

\subsection{Beam Pattern Learning} \label{subsec:simu-BPL}

We first evaluate our proposed DRL-based beam pattern learning solution on learning a single beam that serves a single user with LOS connection to the BS. The selected target user is highlighted in \fref{fig:sce_los} with a red dot.
In \fref{1b-clean}, we compare the performance of the learned single beam with a 32-beam classical beamsteering codebook. As only known, classical beamsteering codebook normally performs very well in LOS scenario. However, our proposed method achieves higher beamforming gain than the best beam in the classical beamsteering codebook, with negligible iterations. More interestingly, with less than $4\times10^4$ iterations, the proposed solution can reach more than $90\%$ of the EGC upper bound. It is worth mentioning that the EGC upper bound can only be reached when the user's channel is known and unquantized phase shifters are deployed. By contrast, our proposed solution can finally achieve almost $95\%$ of the EGC upper bound with 3-bit phase shifters and without any channel information. We also plot the learned beam patterns at three different stages (iteration 1000, 5000, and 100000) during the learning process, which helps understand how the beam pattern evolves over time.
As shown in \fref{1b-clean}, at iteration 1000, the learned beam pattern has very strong side lobes, weakening the main lobe gain to a great extent. At iteration 5000, the gain of the main lobe becomes stronger. However, there are still multiple side lobes with relatively high gains. Finally, at iteration 100000, it can be seen that the main lobe has quite strong gain compared to the other side lobes, having at least 10 dB gain over the second strongest side lobe. And most of the side lobes are below $-20$ dB. Besides, the learned beam pattern captures the EGC beam pattern very well, which explains the good performance it achieves. The slight mismatching is mainly caused by the use of quantized phase shifters, which is with only 3-bit resolution.

\begin{figure}[t]
\centering
  \subfigure[]{\raisebox{5mm}{\includegraphics[width=0.3\linewidth]{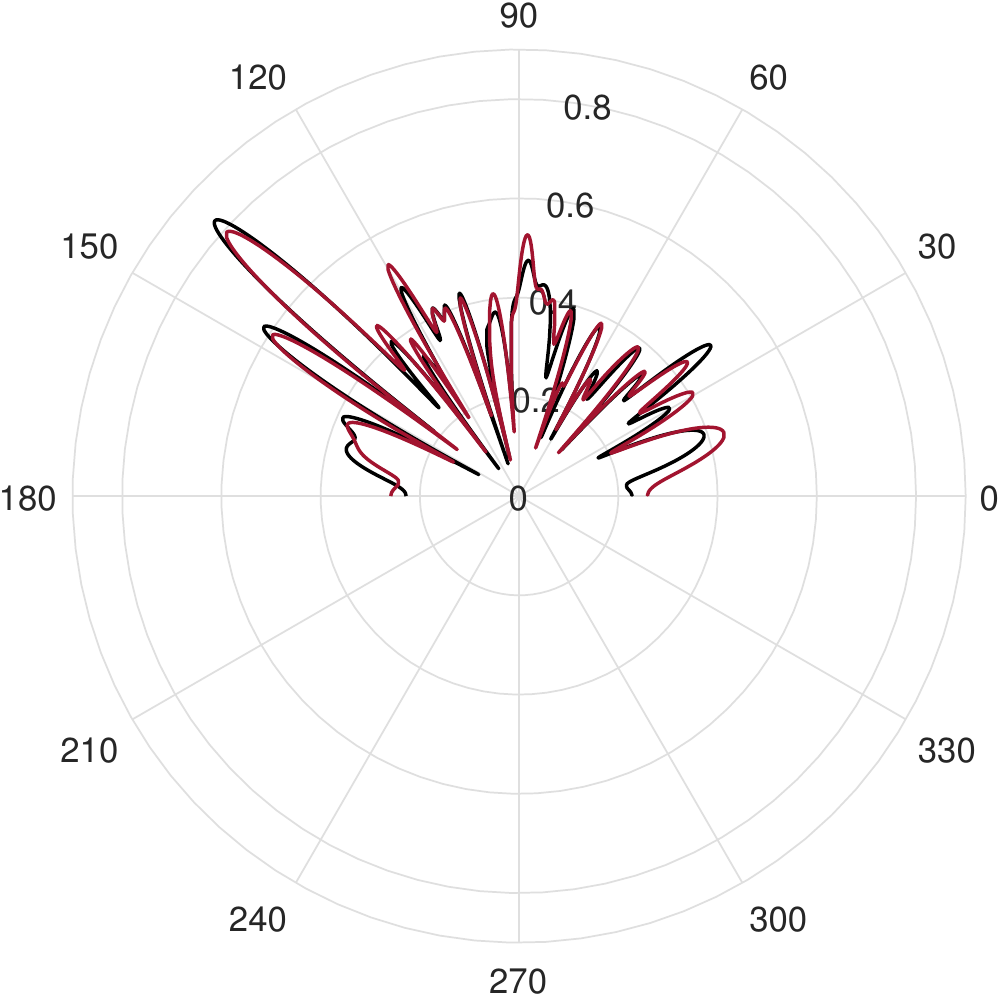} }}
  \subfigure[]{ \includegraphics[width=0.4\linewidth]{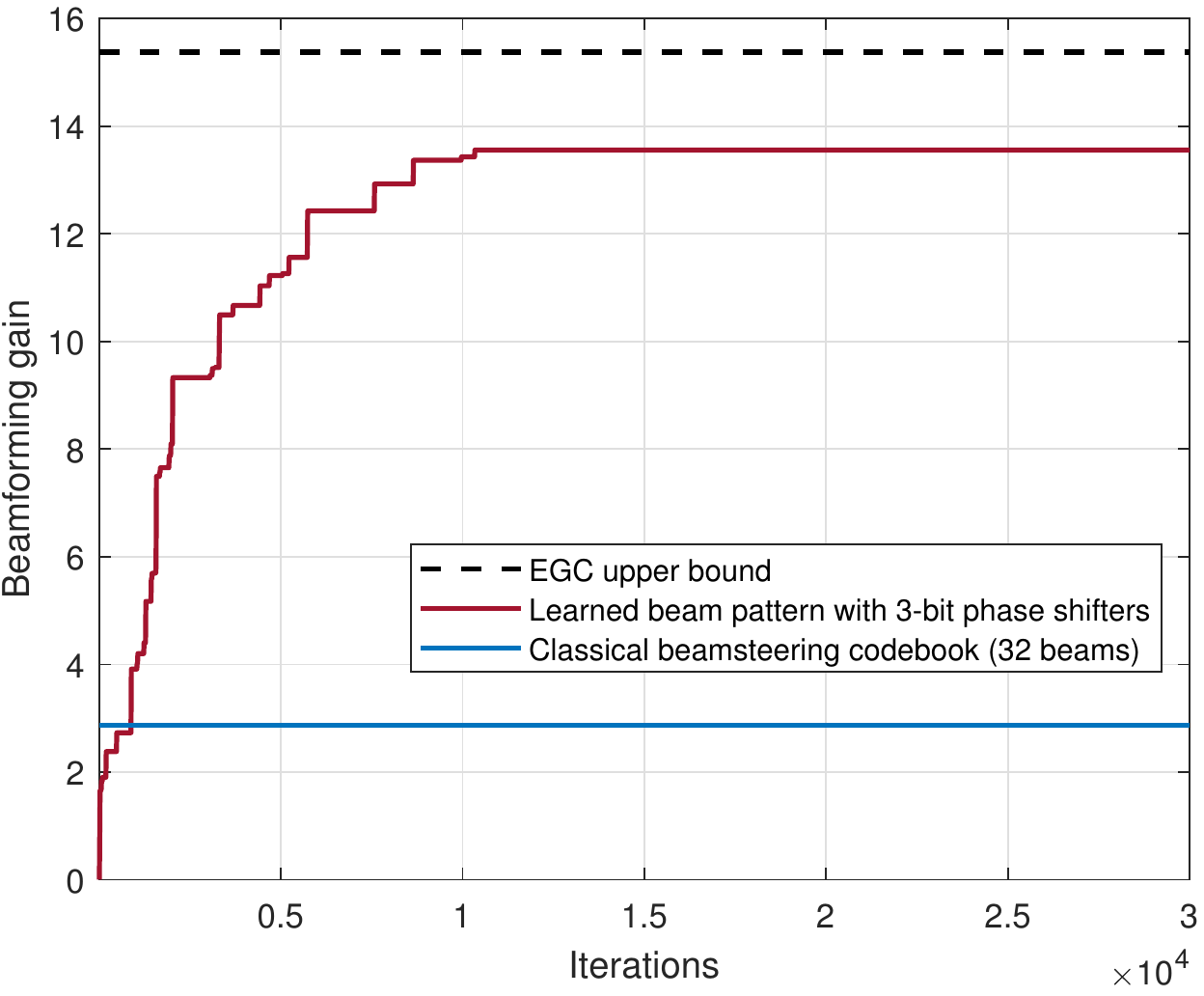} }
  \caption{The beam pattern learned for a single user with LOS connection to the base station. The base station employs a uniform linear array with 32 antennas and 3-bit phase shifters, where hardware impairments exist. The standard deviation of the antenna spacing is $0.1\lambda$ and the standard deviation of the phase mismatches is $0.32\pi$. (a) shows the beam patterns for the equal gain combining/beamforming vector (red) and the learned beam (blue). A transformation of $\sqrt[4]{\cdot}$ is used to better show the finer structure of the beams. (b) shows the learning process.}
  \label{1b-corrupted}
\end{figure}

The proposed beam pattern learning solution is also evaluated on a more realistic situation where hardware impairments exist (with the same user considered above).
The simulation results confirm that our proposed solution is competent to learn optimized beam pattern that adapts to hardware, showing the capability of compensating the unknown hardware mismatches.
\fref{1b-corrupted} (a) shows the beam patterns for both EGC beam and the learned beam. At the first glance, the learned beam appears distorted and has lots of side-lobes. However, the performance of such beam is excellent, which can be explained by comparing its beam pattern with the EGC beam. \textbf{As can be seen from the learned beam pattern, our proposed solution intelligently approximates the optimal beam, where all the dominant lobes are well captured.} By contrast, the classical beamsteering codebook fails when the hardware is not perfect, as depicted in \fref{1b-corrupted} (b). This is because the distorted array pattern incurred by the hardware impairment makes the pointed classical beamsteering beams only able to capture a small portion of the energy, resulting in a huge degradation in beamforming gain. The learned beam shown in \fref{1b-corrupted} (a) is capable of achieving more than $90\%$ of the EGC upper bound with approximately only $10^4$ iterations, as shown in \fref{1b-corrupted} (b). This is especially interesting for the fact that the proposed solution does not rely on any channel state information. As is known, the channel estimation in this case relies first on a full calibration of the hardware, which is a hard and expensive process.

\subsection{Beam Codebook Learning} \label{subsec:simu-BCL}

\begin{figure}[t]
\centering
  \subfigure[]{ \includegraphics[width=0.33\linewidth]{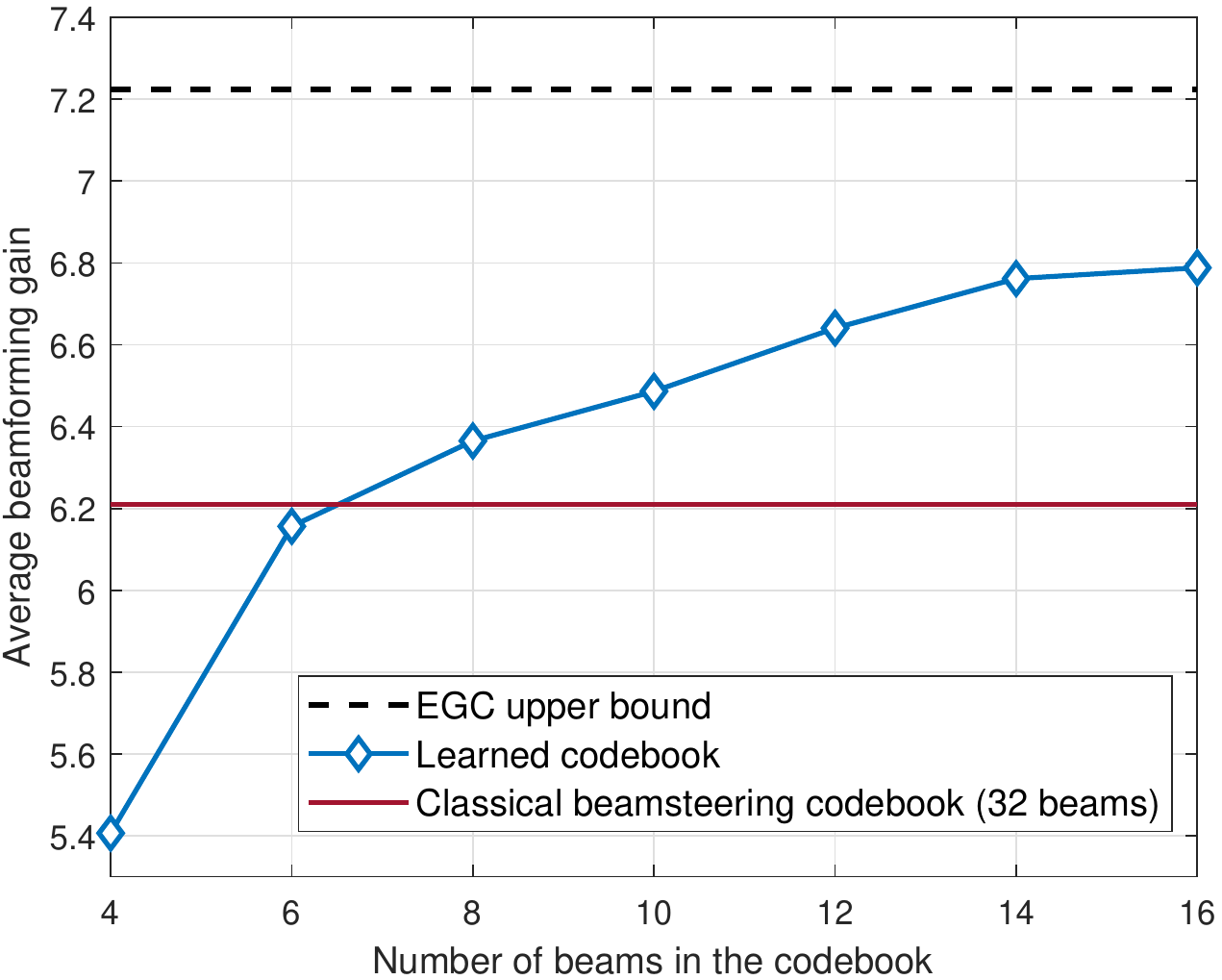} }
  \subfigure[]{ \includegraphics[width=0.33\linewidth]{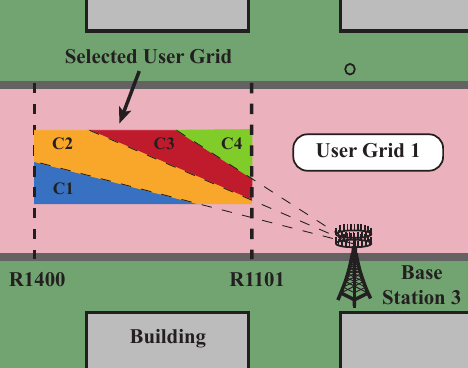} }
  \subfigure[]{ \includegraphics[width=0.27\linewidth]{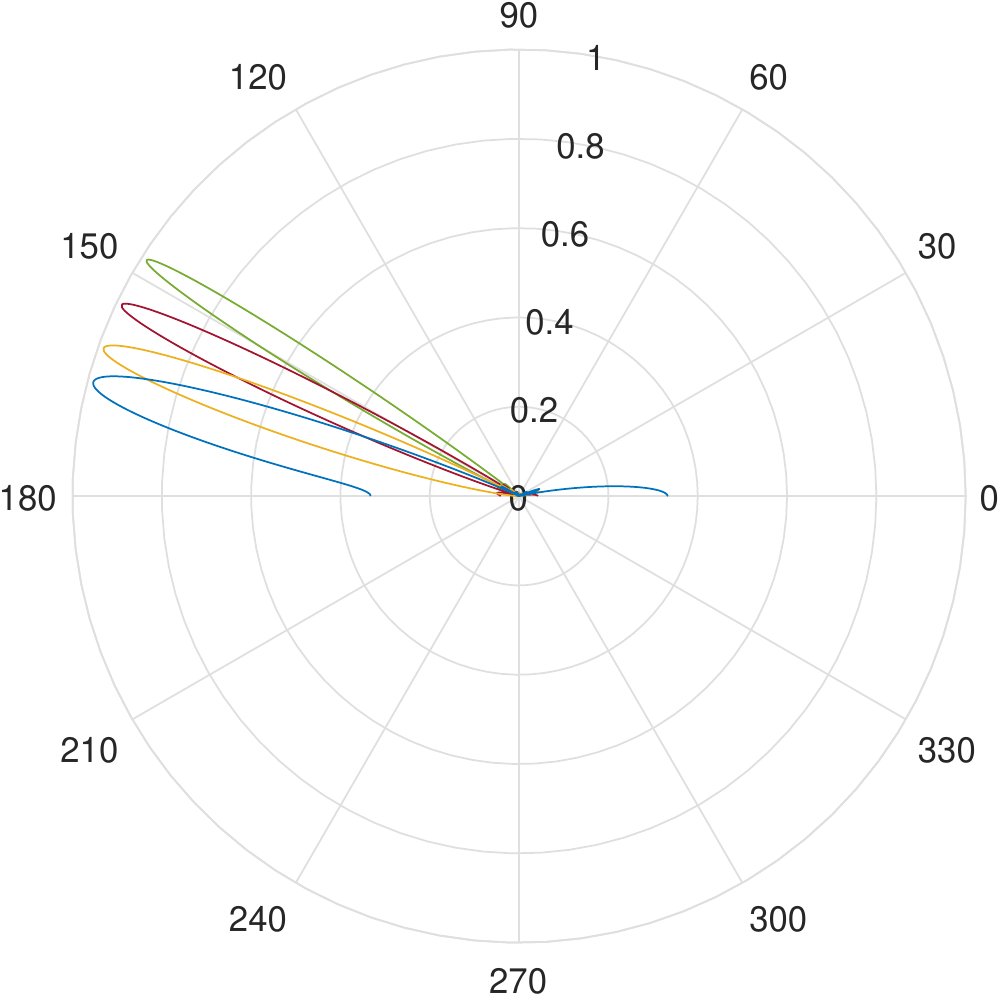} }
  \caption{The learning results of the proposed DRL-based codebook learning solution under a LOS scenario where the base station employs a perfect uniform linear array. (a) shows the average beamforming gain versus the number of beams in the codebook. (b) shows the result of clustering users into 4 groups. (c) shows the beam patterns for the learned 4-beam codebook in (a), which is based on the clustering result in (b).}
  \label{LOS}
\end{figure}

In this subsection, we evaluate our proposed DRL-based beam codebook learning solution in several scenarios. The task of learning a beam codebook with multiple beams is significantly different than learning a single beam (pattern) from computational complexity perspective. For example, for a base station with 32 antennas and 4-bit discrete phase shifters, there are $16^{32}$ possible beamforming vectors, from which a single vector is selected in the beam pattern learning case. However, learning a codebook will further result in finding combinations out of this huge pool. To address this problem, we propose a clustering and assignment approach, given by \aref{alg2}, that essentially decomposes the huge task into $N$ independent, parallel and relatively lightweight sub-tasks. This facilitates the problem of learning a codebook with multiple beams. Before we dive into the discussions, it is important to mention that due to the stationarity of our scenario, clustering/assignment is performed only once in our simulations. If the environment is more dynamic, the clustering/assignment is expected to be done more frequently.

\fref{LOS} (a) plots the average beamforming gain versus the number of beams in the codebook under the LOS scenario shown in \fref{fig:sce_los}, where the BS adopts an ideal uniform linear array. It shows that the average beamforming gain is monotonically increasing as the number of beams increases. Besides, with only 6 beams, the proposed solution has almost the same performance as a 32-beam classical beamsteering codebook. And with 8 beams, it outperforms the 32-beam classical beamsteering codebook. This exhibits how our proposed approach adapts the beams based on the user distributions. \textbf{As a result, it significantly reduces the training overhead by avoiding scanning directions where there is no user at all.} In \fref{LOS} (b), we present the clustering result for the users in this LOS scenario. This is a very important step for learning multiple beams. As stated at the end of \sref{sec:Prob}, the ultimately optimized codebook should have a collection of beams, where each one of them is optimized to serve a group of users with similar channels. The clustering stage is the first step that our proposed solution takes to attain that objective. \fref{LOS} (c) depicts the beam patterns of the learned 4-beam codebook. As shown in the learning result, the proposed solution can cluster users based on the similarity of their channels, and form beams to cover the user grid in order to achieve high beamforming gain.

\begin{figure}[t]
\centering
  \subfigure[]{ \includegraphics[width=0.37\linewidth]{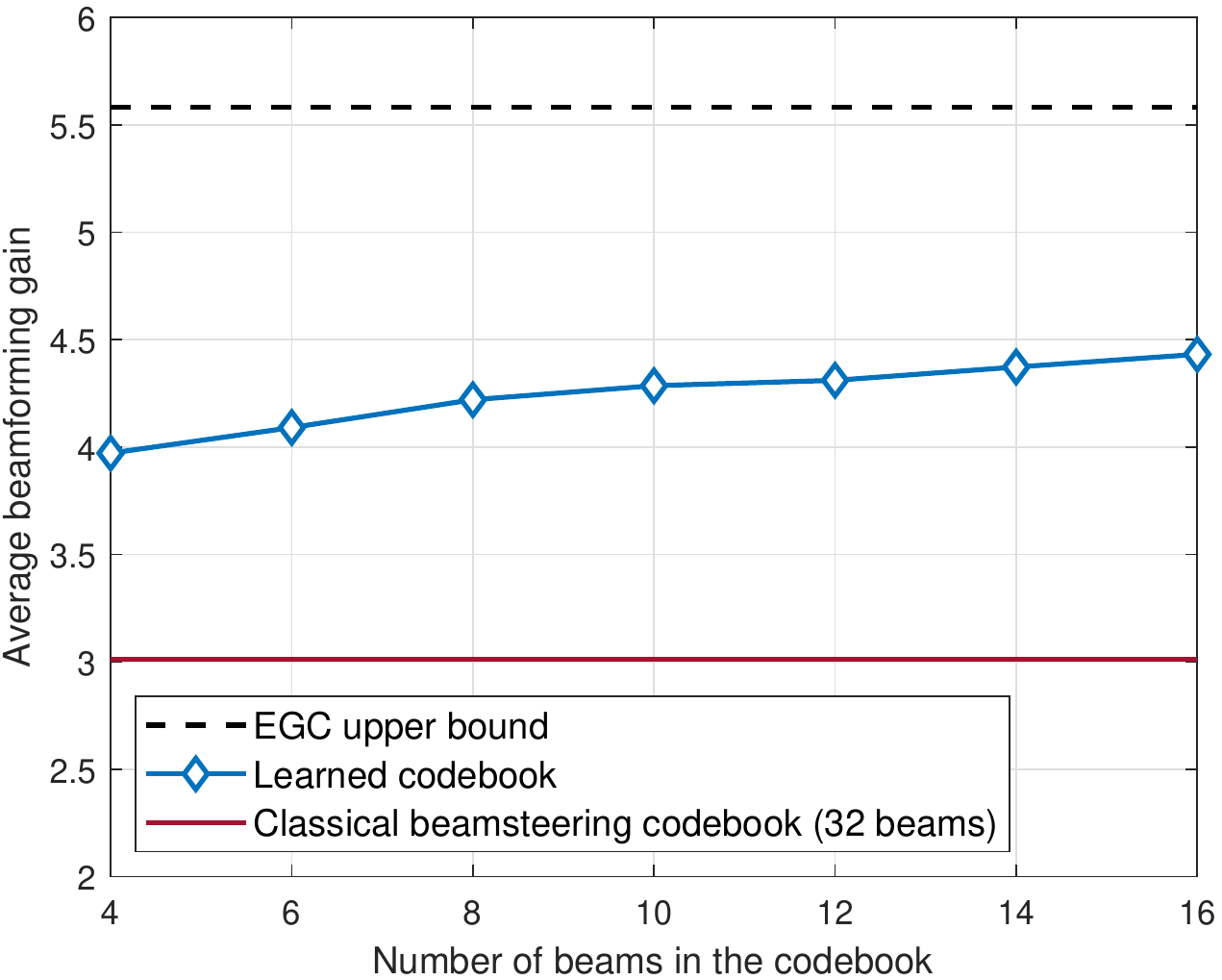} }
  \subfigure[]{ \includegraphics[width=0.58\linewidth]{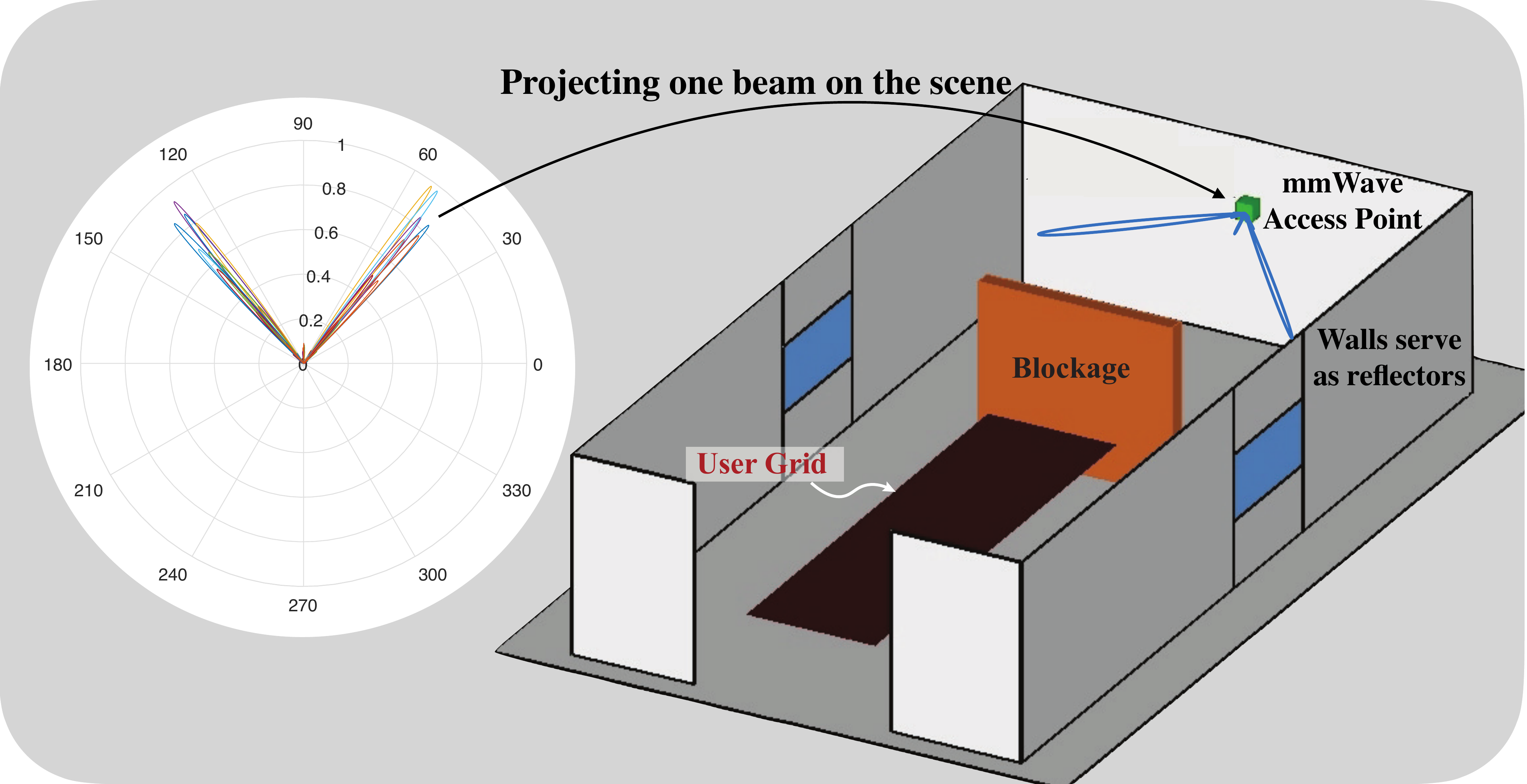} }
  \caption{The learning results of the proposed DRL-based codebook solution under a NLOS scenario. (a) shows the average beamforming gain versus the number of beams in the codebook. (b) shows the beam patterns of the learned 16-beam codebook in (a) and it also shows how one of the learned beams with multi-lobes fits the propagation environment.}
  \label{NLOS}
\end{figure}

We also evaluate the proposed solution under a NLOS scenario shown in \fref{fig:sce_nlos}, where all the users experience NLOS connection with an indoor mmWave access point. As can be seen in \fref{NLOS} (a), the proposed solution surpasses a 32-beam classical beamsteering codebook with only 4 beams. Further, the proposed solution is gradually approaching the EGC upper bound as the size of codebook increases. It should note that in order to achieve the EGC upper bound: (i) The number of beams in the codebook should be equal to the number of users, (ii) continuous phase shifters should be adopted, and most importantly, (iii) accurate channel state information is needed. By contrast, with only 16 beams and 4-bit phase shifters, the proposed solution can reach $80\%$ of the EGC upper bound, relying only on the receive combining gains. \textbf{In other words, our approach not only significantly reduces the beam training overhead, but also avoids the prohibitive cost of estimating the channels.}
To gain more insight, we plot the beam patterns of the learned 16-beam codebook in \fref{NLOS} (b) and project one of the beams on the adopted scene. It can be seen in \fref{NLOS} (b) that the learned beams have multi-lobes, different than the pointed beams learned in the LOS scenario. However, such beams achieve better performance compared to the pointed beamsteering beams. The reason becomes clear when we observe that because of the blockage in the considered scenario, the signals transmitted by the users have to resort to reflections to reach the access point, where the walls at both sides of the room serve as major reflectors. This clearly shows how our proposed solution adapts the beam pattern to the propagation environment, gaining more power by receiving signals from multiple directions.

\begin{figure}[t]
\centering
  \subfigure[]{ \includegraphics[width=0.35\linewidth]{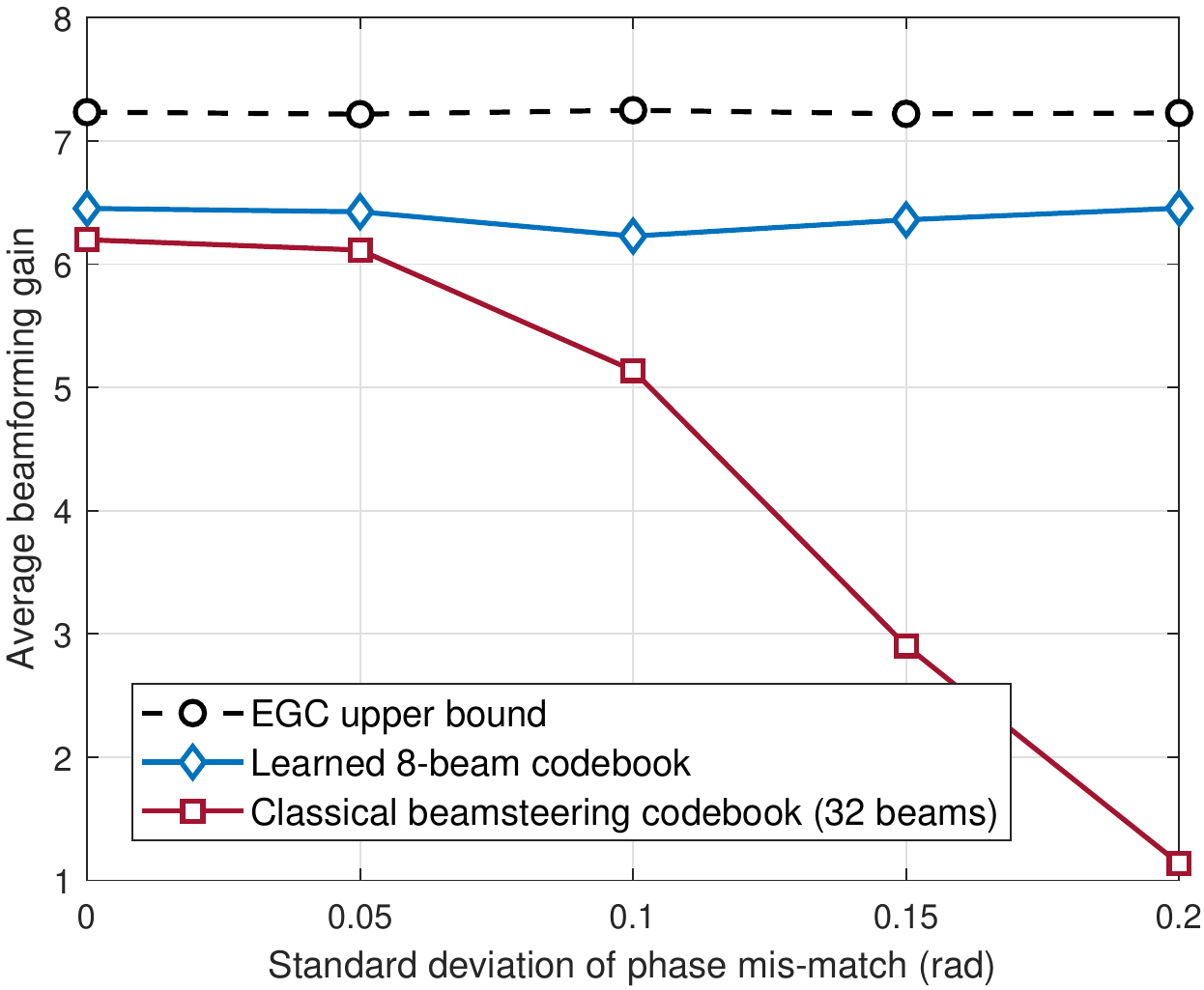} }
  \subfigure[]{ \includegraphics[width=0.29\linewidth]{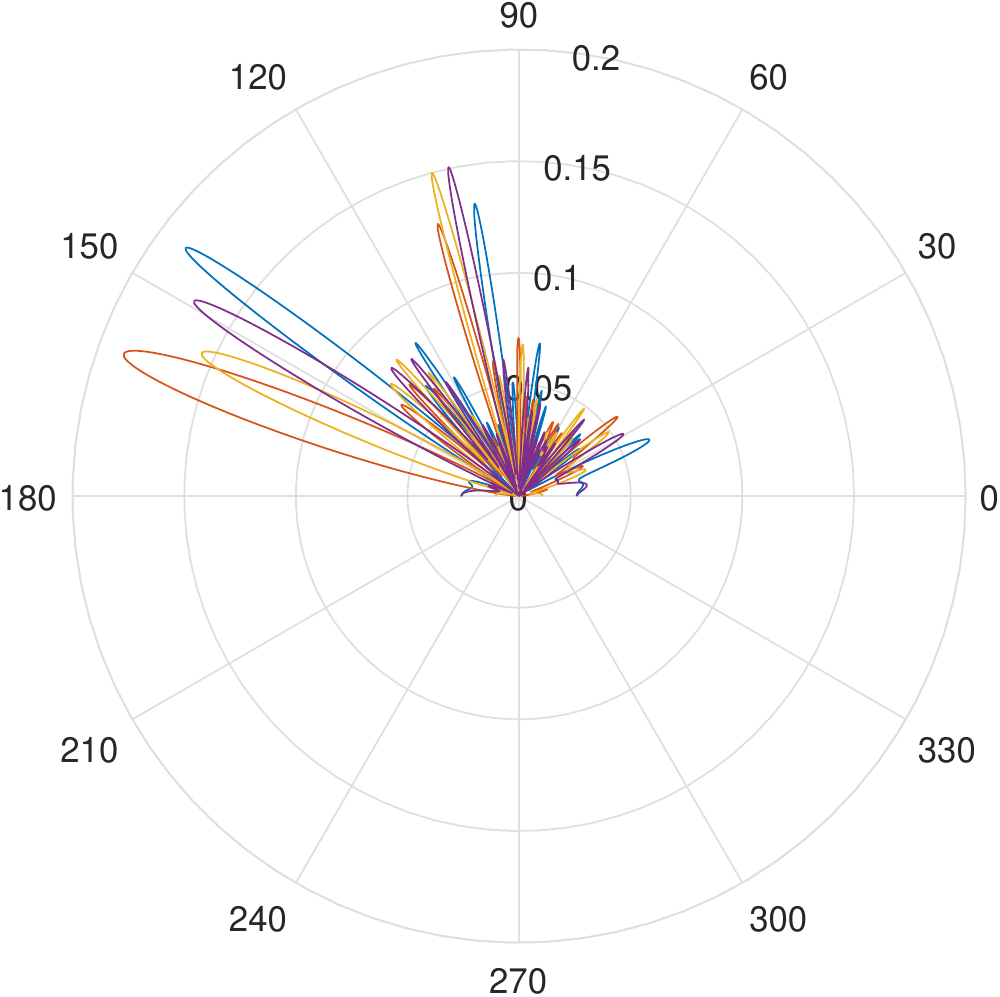} }
  \subfigure[]{ \includegraphics[width=0.29\linewidth]{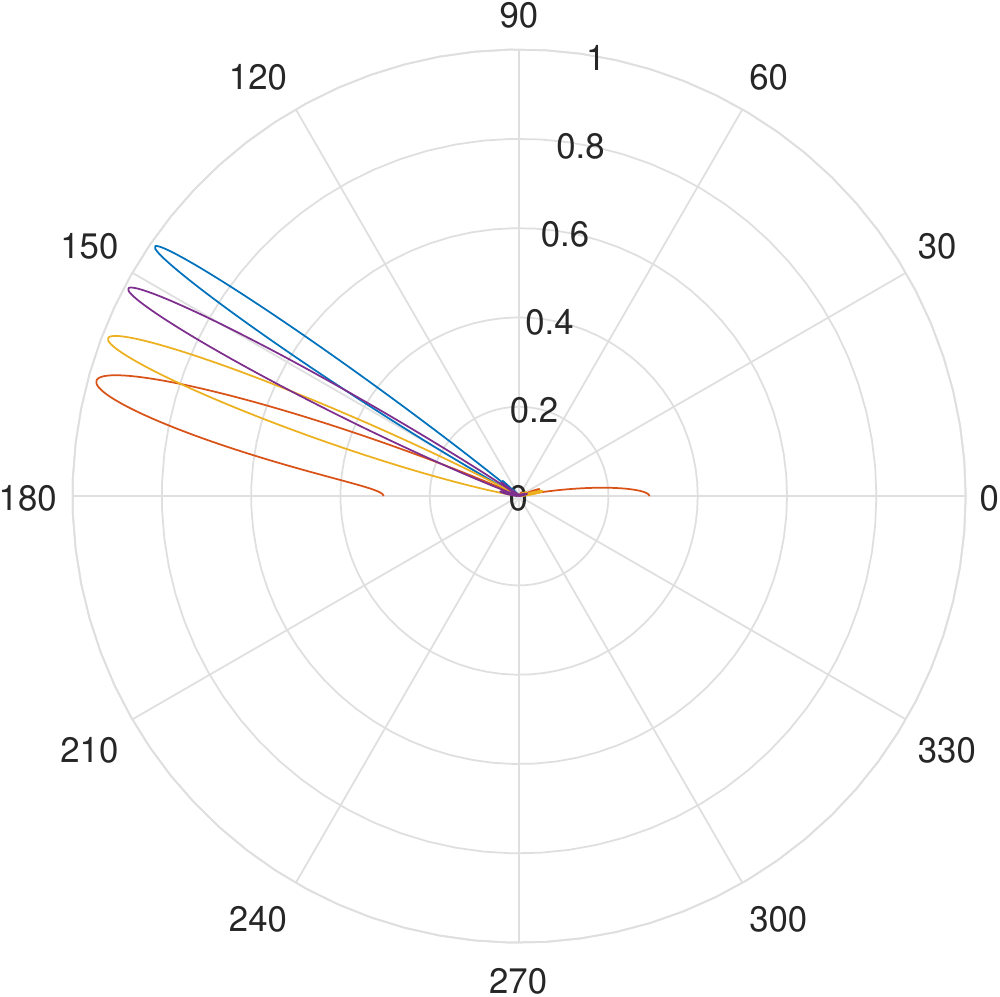} }
  \caption{The learning results of the proposed DRL-based solution under the same LOS scenario with hardware impairments being considered. (a) shows the average beamforming gain versus the standard deviation of phase mismatch, where the antenna spacing mismatch has a fixed standard deviation of $0.1\lambda$. (b) shows the beam patterns of 4 beams in the learned codebook projected onto the ``clean'' angular space. (c) shows the same beams as in (b) projected onto the ``corrupted'' angular space.}
  \label{cLOS}
\end{figure}

Learning codebooks that overcome the hardware impairments is one of the most important application cases of our DRL-based codebook learning approach. Therefore, we evaluate the proposed solution under the same LOS scenario shown in \fref{fig:sce_los}, with hardware impairments being considered. Furthermore, we test our solution under different standard deviations of the phase mismatch, where a fixed antenna spacing mismatch with a standard deviation of $0.1\lambda$ is assumed. For each channel dataset, we learn a 8-beam codebook and compare it with a 32-beam classical beamsteering codebook. In \fref{cLOS} (a), we plot the average beamforming gain versus the standard deviation of the phase mismatch. The result shows that as the standard deviation of the phase mismatch increases, i.e. the hardware impairments become more severe, the proposed DRL based solution keeps a balanced performance. The slight fluctuation is mainly caused by the uncertainty nature of solving the highly non-convex problem \eqref{Prob-0}. By contrast, the performance of the 32-beam classical beamsteering codebook degrades drastically as the level of hardware impairment increases. This empirically shows the robustness of the proposed codebook learning approach to different levels of hardware impairments. In \fref{cLOS} (b), we plot the beam patterns of 4 beams in the learned codebook. It can be seen that these beams have quite distorted beam patterns like the single beam case shown in \fref{1b-corrupted} (a). To show that these distorted beam patterns indeed \textbf{match} the hardware impairments, we plot the same beams in \fref{cLOS} (c), but project them on the ``corrupted'' angular space. As illustrated in \fref{cLOS} (c), the learned beams actually appear ``clean'' and pointy in the corrupted angular space. This empirically verifies the capability of the proposed solution in learning beams that adapt to the flawed hardware.

\section{Conclusions and Discussions} \label{sec:Conclusions}

In this paper, we considered the problem of designing environment and hardware aware beam codebooks for mmWave/THz MIMO systems with hardware-constrained architectures and without requiring channel knowledge. We developed a DRL based framework that learns how to adapt the beam patterns to the surrounding environment, user distribution and hardware impairments, relying only on receive power measurements. We also introduced a clustering-assignment approach to efficiently learn beam codebooks without requiring any knowledge about the user positions and without requiring the users to be stationary during the learning process.  The developed approach is evaluated at various environments, as well as different levels of hardware impairments. Simulation results show promising capability of the proposed solution in learning environment and hardware aware codebooks relying only on the receive power measurements and with finite resolution phase shifters. The learned codebook outperforms the classical beamsteering codebook with much smaller number of beams in all studied scenarios. Further, the  beams learned by the  proposed solution approach the beam shapes of the ideal beams (characterized by unconstrained beamforming vectors under full channel knowledge) and achieve similar SNR performance. For the future work, it will be interesting to extend the developed framework to other MIMO systems that as those adopting hybrid analog/digital architectures.

\linespread{1.1}


\begin{thebibliography}{10}
	\providecommand{\url}[1]{#1}
	\csname url@samestyle\endcsname
	\providecommand{\newblock}{\relax}
	\providecommand{\bibinfo}[2]{#2}
	\providecommand{\BIBentrySTDinterwordspacing}{\spaceskip=0pt\relax}
	\providecommand{\BIBentryALTinterwordstretchfactor}{4}
	\providecommand{\BIBentryALTinterwordspacing}{\spaceskip=\fontdimen2\font plus
		\BIBentryALTinterwordstretchfactor\fontdimen3\font minus
		\fontdimen4\font\relax}
	\providecommand{\BIBforeignlanguage}[2]{{%
			\expandafter\ifx\csname l@#1\endcsname\relax
			\typeout{** WARNING: IEEEtran.bst: No hyphenation pattern has been}%
			\typeout{** loaded for the language `#1'. Using the pattern for}%
			\typeout{** the default language instead.}%
			\else
			\language=\csname l@#1\endcsname
			\fi
			#2}}
	\providecommand{\BIBdecl}{\relax}
	\BIBdecl
	
	\bibitem{zhang2021reinforcement}
	Y.~Zhang, M.~Alrabeiah, and A.~Alkhateeb, ``Reinforcement Learning for Beam
	    Pattern Design in Millimeter Wave and Massive {MIMO} Systems,'' in \emph{Proc.
		of IEEE Asilomar Conference on Signals, Systems, and Computers}, Nov 2020.
	
	\bibitem{Alkhateeb2014MIMO}
	A.~{Alkhateeb}, J.~{Mo}, N.~{Gonzalez-Prelcic}, and R.~W. {Heath}, ``{MIMO
		Precoding and Combining Solutions for Millimeter-Wave Systems},'' \emph{IEEE
		Communications Magazine}, vol.~52, no.~12, pp. 122--131, 2014.
	
	\bibitem{Alkhateeb2014}
	A.~Alkhateeb, O.~El~Ayach, G.~Leus, and R.~Heath, ``{Channel Estimation and
		Hybrid Precoding for Millimeter Wave Cellular Systems},'' \emph{IEEE Journal
		of Selected Topics in Signal Processing}, vol.~8, no.~5, pp. 831--846, Oct.
	2014.
	
	\bibitem{giordani2019}
	M.~{Giordani}, M.~{Polese}, A.~{Roy}, D.~{Castor}, and M.~{Zorzi}, ``{A
		Tutorial on Beam Management for 3GPP NR at mmWave Frequencies},'' \emph{IEEE
		Communications Surveys Tutorials}, vol.~21, no.~1, pp. 173--196, 2019.
	
	\bibitem{11ad}
	\BIBentryALTinterwordspacing
	{IEEE 802.11ad}, ``{IEEE} 802.11ad standard draft {D0.1}.'' [Online].
	Available: \url{www.ieee802.org/11/Reports/tgad update.htm}
	\BIBentrySTDinterwordspacing
	
	\bibitem{Alr_vision}
	M.~{Alrabeiah}, A.~{Hredzak}, and A.~{Alkhateeb}, ``Millimeter wave base
	stations with cameras: Vision-aided beam and blockage prediction,'' in
	\emph{IEEE Vehicular Technology Conference (VTC2020-Spring)}, 2020, pp. 1--5.
	
	
	\bibitem{Lo1999}
	T.~K.~Y. {Lo}, ``Maximum ratio transmission,'' \emph{IEEE Transactions on
		Communications}, vol.~47, no.~10, pp. 1458--1461, 1999.
	
	\bibitem{Love2003}
	D.~Love and R.~Heath~Jr, ``Equal gain transmission in multiple-input
	multiple-output wireless systems,'' \emph{IEEE Transactions on
		Communications}, vol.~51, no.~7, pp. 1102--1110, 2003.
	
	\bibitem{Li2017}
	X.~{Li}, Y.~{Zhu}, and P.~{Xia}, ``{Enhanced Analog Beamforming for Single
		Carrier Millimeter Wave MIMO Systems},'' \emph{IEEE Transactions on Wireless
		Communications}, vol.~16, no.~7, pp. 4261--4274, 2017.
	
	\bibitem{ElAyach2014}
	O.~El~Ayach, S.~Rajagopal, S.~Abu-Surra, Z.~Pi, and R.~Heath, ``Spatially
	sparse precoding in millimeter wave {MIMO} systems,'' \emph{IEEE Transactions
		on Wireless Communications}, vol.~13, no.~3, pp. 1499--1513, Mar. 2014.
	
	\bibitem{Hur2013}
	S.~Hur, T.~Kim, D.~Love, J.~Krogmeier, T.~Thomas, and A.~Ghosh, ``Millimeter
	wave beamforming for wireless backhaul and access in small cell networks,''
	\emph{IEEE Transactions on Communications}, vol.~61, no.~10, pp. 4391--4403,
	Oct. 2013.
	
	\bibitem{Wang2009}
	J.~Wang,  \emph{et~al.}, ``Beam codebook based beamforming
	protocol for multi-{Gbps} millimeter-wave {WPAN} systems,'' \emph{IEEE
		Journal on Selected Areas in Communications}, vol.~27, no.~8, pp. 1390--1399,
	Nov. 2009.
	
	\bibitem{Alkhateeb2016d}
	A.~Alkhateeb and R.~W. Heath, ``Frequency selective hybrid precoding for
	limited feedback millimeter wave systems,'' \emph{IEEE Transactions on
		Communications}, vol.~64, no.~5, pp. 1801--1818, May 2016.
	
	\bibitem{Alrabeiah2020Neural}
	M.~Alrabeiah, Y.~Zhang, and A.~Alkhateeb, ``{Neural Networks Based Beam
		Codebooks: Learning mmWave Massive MIMO Beams that Adapt to Deployment and
		Hardware},'' 2020.
	
	\bibitem{Dulacarnold2015}
	G.~Dulac-Arnold, R.~Evans, H.~van Hasselt, P.~Sunehag, T.~Lillicrap, J.~Hunt,
	T.~Mann, T.~Weber, T.~Degris, and B.~Coppin, ``Deep reinforcement learning in
	large discrete action spaces,'' 2015.
	
	\bibitem{DeepMIMO}
	A.~Alkhateeb, ``{Deep{MIMO}: A Generic Deep Learning Dataset for Millimeter
		Wave and Massive {MIMO} Applications},'' in \emph{Proc. of Information Theory
		and Applications Workshop (ITA)}, San Diego, CA, Feb 2019, pp. 1--8.
	
	\bibitem{HeathJr2016}
	R.~W. Heath, \emph{et~al.} ``An
	overview of signal processing techniques for millimeter wave {MIMO}
	systems,'' \emph{IEEE Journal of Selected Topics in Signal Processing},
	vol.~10, no.~3, pp. 436--453, April 2016.
	
	\bibitem{Alrabeiah2019}
	M.~{Alrabeiah} and A.~{Alkhateeb}, ``{Deep Learning for TDD and FDD Massive
		MIMO: Mapping Channels in Space and Frequency},'' \emph{arXiv e-prints}, p.
	arXiv:1905.03761, May 2019.
	
	\bibitem{Pal2010}
	P.~{Pal} and P.~P. {Vaidyanathan}, ``Nested arrays: A novel approach to array
	processing with enhanced degrees of freedom,'' \emph{IEEE Transactions on
		Signal Processing}, vol.~58, no.~8, pp. 4167--4181, Aug 2010.
	
	\bibitem{Rubsamen2009}
	M.~{Rubsamen} and A.~B. {Gershman}, ``Direction-of-arrival estimation for
	nonuniform sensor arrays: From manifold separation to fourier domain music
	methods,'' \emph{IEEE Transactions on Signal Processing}, vol.~57, no.~2, pp.
	588--599, 2009.
	
	\bibitem{Moon2019}
	T.~{Moon}, J.~{Gaun}, and H.~{Hassanieh}, ``Online millimeter wave phased array
	calibration based on channel estimation,'' in \emph{2019 IEEE 37th VLSI Test
		Symposium (VTS)}, 2019, pp. 1--6.
	
	\bibitem{Mnih2013}
	V.~Mnih, K.~Kavukcuoglu, D.~Silver, A.~Graves, I.~Antonoglou, D.~Wierstra, and
	M.~Riedmiller, ``{Playing Atari with Deep Reinforcement Learning},'' 2013.
	
	\bibitem{Mnih2015}
	V.~Mnih \emph{et~al.}, ``{Human-level Control through Deep Reinforcement
		Learning},'' \emph{Nature}, vol. 518, no. 7540, pp. 529--533, 2015.
	
	\bibitem{Sutton2018}
	R.~S. Sutton and A.~G. Barto, \emph{Reinforcement Learning: An
		Introduction}.\hskip 1em plus 0.5em minus 0.4em\relax Cambridge, MA, USA: A
	Bradford Book, 2018.
	
	\bibitem{Timothy2015}
	T.~P. Lillicrap, J.~J. Hunt, A.~Pritzel, N.~Heess, T.~Erez, Y.~Tassa,
	D.~Silver, and D.~Wierstra, ``Continuous control with deep reinforcement
	learning,'' 2015.
	
	\bibitem{Uhlenbeck1930}
	\BIBentryALTinterwordspacing
	G.~E. Uhlenbeck and L.~S. Ornstein, ``On the Theory of the Brownian Motion,''
	\emph{Phys. Rev.}, vol.~36, pp. 823--841, Sep 1930. [Online]. Available:
	\url{https://link.aps.org/doi/10.1103/PhysRev.36.823}
	\BIBentrySTDinterwordspacing
	
	\bibitem{Zhang2020}
	Y.~Zhang, M.~Alrabeiah, and A.~Alkhateeb, ``{Learning Beam Codebooks with
		Neural Networks: Towards Environment-Aware mmWave MIMO},'' 2020.
	
	\bibitem{PatternRecog}
	C.~M. Bishop, \emph{Pattern recognition and machine learning}.\hskip 1em plus
	0.5em minus 0.4em\relax springer, 2006.
	
	\bibitem{HungarianAlgo}
	H.~W. Kuhn, ``The Hungarian method for the assignment problem,'' \emph{Naval
		research logistics quarterly}, vol.~2, no. 1-2, pp. 83--97, 1955.
	
	\bibitem{EffBackProp}
	Y.~A. LeCun, L.~Bottou, G.~B. Orr, and K.-R. M{\"u}ller, ``Efficient
	backprop,'' in \emph{Neural networks: Tricks of the trade}.\hskip 1em plus
	0.5em minus 0.4em\relax Springer, 2012, pp. 9--48.
	
	\bibitem{Zhang2020Deep}
	Y.~{Zhang}, M.~{Alrabeiah}, and A.~{Alkhateeb}, ``{Deep Learning for Massive
		MIMO with 1-Bit ADCs: When More Antennas Need Fewer Pilots},'' \emph{IEEE
		Wireless Communications Letters}, 2020.
	
	\bibitem{Alkhateeb2015Limited}
	A.~{Alkhateeb}, G.~{Leus}, and R.~W. {Heath}, ``Limited feedback hybrid
	precoding for multi-user millimeter wave systems,'' \emph{IEEE Transactions
		on Wireless Communications}, vol.~14, no.~11, pp. 6481--6494, 2015.
	
\end{thebibliography}
\end{document}